\documentclass[prd,aps,nofootinbib,twocolumn]{revtex4}
\usepackage{amssymb,amsmath,epsfig}
\usepackage[bookmarks]{hyperref}
\usepackage[usenames,dvipsnames]{color}
\usepackage{multirow}
\usepackage{soul}

\newcommand{\Complex}{\mathbb{C}}
\newcommand{\diag}{\mbox{diag}}

\newcommand{\proof}{\noindent {\bf Proof. }}

\newcommand{\qed}{\hfill \fbox{} \vspace{.3cm}}

\newtheorem{theorem}{Theorem}

\newcommand{\eq}[1]{Eq.\,\eqref{#1}}
\newcommand{\eqs}[1]{Eqs.\,\eqref{#1}}
\newcommand{\ud}{\mathrm{d}}
\newcommand{\co}{\mathrm{const.}}

\begin{document}

\title{Choked accretion onto a Kerr black hole}
\author{Alejandro Aguayo-Ortiz\footnote{Email: aaguayo@astro.unam.mx}$^1$,
Olivier Sarbach$^2$ and Emilio Tejeda$^3$}

\affiliation{$^1$ Instituto de Astronom\'ia, Universidad Nacional Aut\'onoma
de M\'exico, AP 70-264, 04510 Ciudad de M\'exico, M\'exico,}

\affiliation{$^2$Instituto de F\'\i sica y Matem\'aticas,
Universidad Michoacana de San Nicol\'as de Hidalgo,
Edificio C-3, Ciudad Universitaria, 58040 Morelia, Michoac\'an, M\'exico,}

\affiliation{$^3$C\'atedras Conacyt --  Instituto de F\'\i sica y Matem\'aticas,
Universidad Michoacana de San Nicol\'as de Hidalgo,
Edificio C-3, Ciudad Universitaria, 58040 Morelia, Michoac\'an, M\'exico.} 


\begin{abstract}
The choked accretion model consists of a purely hydrodynamical mechanism
in which, by setting an equatorial to polar density contrast, a spherically
symmetric accretion flow transitions to an inflow-outflow configuration. This
scenario has been studied in the case of a (non-rotating) Schwarzschild black
hole as central accretor, as well as in the non-relativistic limit. In this
article, we generalize these previous works by studying the accretion of
a perfect fluid onto a (rotating) Kerr black hole. We first describe the
mechanism by using a steady-state, irrotational analytic solution of an
ultrarelativistic perfect fluid, obeying a stiff equation of state. We then
use hydrodynamical numerical simulations in order to explore a more general
equation of state. Analyzing the effects of the black hole's rotation on
the flow, we find in particular that the choked accretion inflow-outflow
morphology prevails for all possible values of the black hole's spin parameter,
showing the robustness of the model.
\end{abstract}


\date{\today}


\maketitle

\section{Introduction}
\label{Sec:introduction}

Black hole accretion theory has been an important building-block of our
current understanding of high-energy astrophysical phenomena such as X-ray
Binaries, Gamma Ray Bursts, and Active Galactic Nuclei~\cite{mAcF2013}. In
recent years, this field of knowledge has gone through a revolution led
by the observational breakthrough of gravitational wave astronomy, that
has allowed a systematic analysis of close to fifty binary black hole mergers
to date~\cite{ligo2019,ligo2020}, as well as the extreme-resolution imaging of the
immediate environment of astrophysical black holes achieved by the Event
Horizon Telescope~\cite{EHT_M87}.

Since the pioneering work of Bondi~\cite{hB52}, the introduction of exact,
analytic solutions for modeling different astrophysical scenarios has been
instrumental in the continuous development of accretion theory. Analytic
models, by transparently highlighting the role played by different physical
ingredients, are key in cementing our understanding and building our intuition
around the studied phenomena. Moreover, analytic solutions are crucial tools
as benchmark tests for numerical codes~\cite{RZ13}.

Within the regime of Newtonian gravity, the Bondi solution describes the
stationary flow of a spherically symmetric gas cloud accreting onto a
gravitational object. This solution was extended by Michel~\cite{fM72}
to a relativistic regime by considering a Schwarzschild black hole as
central accretor. On the other hand, analytic solutions to the so-called
wind accretion scenario have been introduced by Bondi and Hoyle~\cite{hoyle}
and Hoyle and Littleton~\cite{hoylely} in the Newtonian context as well as
by Tejeda and Aguayo-Ortiz~\cite{eTaA19} in the relativistic context of a
Schwarzschild black hole.\footnote{Also see~\cite{tejeda1,tejeda2,tejeda3}
for a related, analytic model of a rotating dust cloud accreting onto a
rotating or a non-rotating black hole.} Several analytic and numerical
investigations have further extended the study of spherical accretion,
e.g.~\cite{Nobili91,Karkowski06,pMeM08,Fragile12,Roedig12,Sadowski13,McKinney14,fLmGfG14,eCoS15a,eCmMoS15,Weih20},
as well as of wind accretion,
e.g.~\cite{Hunt71,Shima85,font98b,Ruffert94,font99,LG13}.

It is important to note, however, that although astrophysical black holes are
expected to rotate in general, very few analytic solutions exist for rotating
black holes as described by the Kerr metric. A notable exception is the
analytic solution introduced by Petrich, Shapiro and Teukolsky~\cite{lPsSsT88}
that describes, under the assumptions of steady-state and irrotational flow,
an ultrarelativistic stiff fluid accreting onto a Kerr black hole.

Based on the general solution presented in~\cite{lPsSsT88}, and following
on the work of~\cite{hernandez14} and~\cite{aAeTxH19}, Tejeda, Aguayo-Ortiz
and Hernandez~\cite{TAH20} presented a simple, hydrodynamical mechanism for
launching bipolar outflows from a choked accretion flow onto a Schwarzschild
black hole.
This model starts from a spherically symmetric accretion flow onto a central
massive object and introduces a deviation away from spherical symmetry
by considering a small-amplitude, large-scale density gradient in such a
way that the equatorial region of the accreting material is over dense as
compared to the polar regions.
This anisotropic density field  translates into a pressure-driven force that,
provided a sufficiently large mass accretion rate, can deflect a fraction of
the originally radial accretion flow onto a bipolar outflow. The threshold
value for the accretion rate determining whether the inflow chokes and the
launching mechanism is successful or not is found to be of the order of
the mass accretion rate corresponding to the spherically symmetric cases
discussed by Bondi and Michel.

Even though the approximation of a stiff fluid has a rather limited
applicability in astrophysics, the mechanism presented in~\cite{TAH20}
was shown to be valid for more general equations of state by means
of full hydrodynamic numerical simulations. Moreover, as discussed
in~\cite{aAeTxH19}, this mechanism is also valid in the context of Newtonian
gravity.

In this work, we present an extension of the choked accretion model introduced
in~\cite{aAeTxH19,TAH20} to the case of a rotating central black hole as
described by the Kerr metric. We study this problem using both an analytic
solution for an ultrarelativistic stiff fluid as well as full hydrodynamic
numerical simulations for fluids described by more general equations of
state. In addition to demonstrating that the choked accretion mechanism can
successfully operate with a central rotating black hole, we also analyze
the effects of the black hole rotation on the accretion flow.

We focus mostly on the case in which the axis of the bipolar outflow
is aligned with the black hole's rotation axis, although we also briefly
discuss the case of a possible misalignment between these two. Considering
that the infalling gas might come from the inner edge of an accretion disk,
we believe that the restriction of alignment is well justified in view of
the Bardeen-Petterson effect~\cite{BP75,LTI19}, which foresees that the inner
part of an accretion disk around a rotating black hole will be aligned with
the equatorial plane of the central black hole.

The choked accretion mechanism introduced in~\cite{aAeTxH19,TAH20} can be
considered as a hydrodynamic toy model of the central engine in
astrophysical scenarios involving both equatorial accretion flows and bipolar
outflows. These scenarios can range from the jets and winds associated with
some Young Stellar Objects to the accretion disk-jet systems associated
with stellar mass black holes (such as X-Ray Binaries and Gamma-Ray Bursts)
as well as with supermassive black holes (such as Radio Loud Galaxies and
other Active Galactic Nuclei).

Even though the choked accretion model does not account directly for fluid
rotation, the assumption of an anisotropic density field is motivated precisely
as a way to introduce indirectly one of the effects of fluid rotation and
angular momentum conservation, namely, the existence of a well-defined
symmetry axis (the rotation axis) and the accompanying flattening of the
accreting fluid that results in an equator-to-poles density gradient (see,
e.g.~\citep{proga2003,mach2018}).

Several works in the literature have studied before different accretion
scenarios featuring both equatorial inflows and bipolar outflows, particularly
within the regime of Hot Accretion Flows \citep{SLE76,YuanNarayan14},
that correspond to geometrically thick, optically thin, and radiatively
inefficient accretion flows. These studies have been both analytic,
with models such as Advection Dominated Accretion Flows (ADAF)
\citep{ADAF94,ADAF95} or Adiabatic Inflow-Outflow Solutions (ADIOS)
\cite{ADIOS99,ADIOS04,ADIOS12}, as well as based on numerical simulations
\citep{INA03,HawleyKrolik06,Proga07,Tchekhovskoy+11,Narayan+12,Waters+20}.
From the point of view of the incorporated physical ingredients, these models
are more realistic than the choked accretion scenario discussed here as they
account for effects such as fluid rotation, viscous dissipation of energy and
transport of angular momentum, interaction with a radiation field, magnetic
fields, among others. Nevertheless, we believe that, given its simplicity and
reliance on pure hydrodynamics, the choked accretion mechanism might be already
at work in some of those systems, acting alongside more complex processes.

Also note that the choked accretion model shares some broad, qualitative
features with some versions of Hot Accretion Flows \citep{YuanNarayan14},
namely, an accreting, quasi-spherical gas distribution, with sub-Keplerian
rotation, and with such a large internal energy that parcels of it become
gravitationally unbound from the central object and can be ejected as
bipolar outflows.

The paper is organized as follows. Based on the approximations of steady-state
and irrotational flow, in Sec.~\ref{Sec:SteadyStateKerr} we present the general
solution of an ultrarelativistic stiff fluid in Kerr spacetime. In contrast
to~\cite{lPsSsT88}, who adopted the Boyer-Lindquist coordinates for this
derivation, we shall employ horizon-penetrating coordinates which are regular
across the black hole's event horizon and allow for a clearer and, in fact,
simpler derivation of the solution.  In Sec.~\ref{Sec:AxisymmetricQuadFlow} we
restrict our discussion on the axisymmetric, quadrupolar solution and discuss
its most salient properties. Based on this solution, in Sec.~\ref{Sec:choked}
we introduce and discuss the analytic model describing choked accretion in a
Kerr spacetime. In Sec.~\ref{Sec:Numeric} we complement this study by means
of hydrodynamic simulations for a more general equation of state. Finally,
in Sec.~\ref{Sec:Conclusions}, we present a summary of the model and give
our conclusions.  Technical details regarding the region of validity of the
analytic solution, a non-axisymmetric exact solution, and convergence
tests of our numerical results are discussed in three appendices. Throughout
this article we use the signature convention $(-,+,+,+)$ for the spacetime
metric and work in geometrized units for which $G=c=1$.

\section{Steady-state, irrotational solutions for an ultrarelativistic
stiff equation of state on a Kerr background spacetime}

\label{Sec:SteadyStateKerr}

In this section, we review the analytic approach of \cite{lPsSsT88} for
obtaining exact, irrotational, steady-state solutions of the relativistic
Euler equations on a Kerr black hole background with an ultrarelativistic
stiff equation of state. We start in Sec.~\ref{SubSec:Derivation} with the
derivation of the Petrich-Shapiro-Teukolsky solution \cite{lPsSsT88} in horizon-penetrating
coordinates. Next, in Sec.~\ref{SubSec:ZAMO} we compute the components of the
three-velocity of the fluid with respect to the reference frame associated with
zero angular momentum observers (ZAMOs), which are naturally adapted to
the Killing symmetries of the Kerr geometry and reduce to the usual static
observers in the non-rotating limit. Finally, in Sec.~\ref{SubSec:Conservation}
we discuss the conserved quantities obeyed by the fluid field, such as the
(rest) mass and energy accretion rates which are important for the physical
interpretation of our model, as well as the angular momentum accretion rate.

An ultrarelativistic stiff equation of state is characterized by the fluid's
pressure $P = K \rho^2$ being proportional to the square of the rest-mass
density $\rho$ and the internal energy dominating the rest mass energy. For
a perfect fluid in local thermodynamical equilibrium obeying the first law
$\ud h = \ud P/\rho$, this implies that its specific enthalpy $h = 2K \rho$
is proportional to $\rho$. Together with the irrotational condition such a
fluid can be described by a scalar potential $\Phi$ satisfying the linear
wave equation
\begin{equation}
\nabla^\mu\nabla_\mu\Phi 
 =  \frac{1}{\sqrt{-g}}\partial_\mu(\sqrt{-g} g^{\mu\nu}\partial_\nu\Phi ) = 0.
\label{Eq:Wave}
\end{equation}
The potential $\Phi$ determines the fluid's specific enthalpy and four-velocity
$U^\mu$ according to
\begin{equation}
h = \sqrt{-(\nabla^\mu\Phi)(\nabla_\mu\Phi)},\qquad
U^\mu = \frac{1}{h}\nabla^\mu\Phi,
\label{Eq:nU}
\end{equation}
from which the rest-mass density and the pressure can also be obtained. An
important point to notice is that not every solution of the wave
equation~(\ref{Eq:Wave}) yields a valid solution for the fluid; indeed,
for $h$ to be well-defined the gradient $\nabla^\mu\Phi$ of $\Phi$ needs to
be timelike.

The key observation in~\cite{lPsSsT88} is that for a steady-state
configuration on a Kerr background, \eq{Eq:Wave} can be decoupled into
standard spherical harmonics (even though the Kerr spacetime is not
spherically symmetric!), leading to a general solution which can be
expressed in terms of well-known special functions. In the following,
we briefly repeat the arguments leading to this expression. However,
unlike the Boyer-Lindquist coordinates used in~\cite{lPsSsT88}, we base
our calculations on the Kerr-type coordinates\footnote{These coordinates
are related to the Kerr coordinates $(v,\phi,r,\theta)$ found in standard
textbooks~\cite{MTW-Book,HawkingEllis-Book} by the transformation $v = t +
r$, and they are related to the standard Boyer-Lindquist coordinates $(t_{\rm
BL},\phi_{\rm BL},r_{\rm BL},\theta_{\rm BL})$ through the transformation
\mbox{$r = r_{\rm BL}$}, \mbox{$\theta = \theta_{\rm BL}$}, while
\begin{subequations}
\begin{eqnarray}
t &=& t_{\rm BL} + \frac{2M}{r_+ - r_-}
 \left[ r_+\ln\left( \frac{r - r_+}{r_+} \right) - r_-\ln\left( \frac{r - r_-}{r_-} \right) \right],
\label{Eq:tBL}\\
\phi &=& \phi_{\rm BL} + \frac{a}{r_+ - r_-}\ln\left( \frac{r - r_+}{r - r_-} \right).
 \label{Eq:phiBL}
\end{eqnarray}
\end{subequations}
} $(t,\phi,r,\theta)$. This has at least two advantages. First, as we
will see, the derivation and final expression for the analytic solution of
\eq{Eq:Wave} is clearer and simpler in terms of these coordinates. Second,
and most importantly, it greatly facilitates the understanding of the
properties of the flow at the horizon, since these coordinates cover the
(future) event horizon $r = r_+$ in addition to the outside region $r > r_+$
(whereas the Boyer-Lindquist coordinates are ill-defined at the horizon).

\subsection{Derivation of the Petrich-Shapiro-Teukolsky solution in the
Kerr-type coordinates}
\label{SubSec:Derivation}

In terms of the coordinates $(t,\phi,r,\theta)$, the Kerr metric components
have determinant $g := \det(g_{\mu\nu}) = -\varrho^4\sin^2\theta$ and the
components of the inverse metric are
\begin{equation}
(g^{\mu\nu} ) = \frac{1}{\varrho^2}\left( \begin{array}{cccc}
-(\varrho^2 + 2M r) & 0 & 2M r & 0 \\
0 & \frac{1}{\sin^2\theta} & a & 0 \\
2M r & a & \Delta & 0 \\
0 & 0 & 0 & 1
\end{array} \right),
\end{equation}
where we use the standard abbreviations\footnote{ We warn the reader that
throughout this work we follow the convention of Ref.~\cite{mAcF2013}
where the similar-looking symbols $\rho$ and $\varrho$ denote the rest-mass
density and the metric coefficient $\varrho = \sqrt{r^2 + a^2\cos^2\theta
}$, respectively.}
$$
\varrho^2 = r^2 + a^2\cos^2\theta,\quad \Delta = r^2 - 2M r + a^2.
$$
Here, $M$ and $a$ are the mass and rotation parameter of the Kerr spacetime,
and we assume that $a^2 < M^2$ such that this spacetime describes a
non-extremal black hole with angular momentum $J = a M$.

With these coordinates, the wave equation~(\ref{Eq:Wave}) assumes the
following explicit form:
\begin{eqnarray}
&& (\varrho^2 + 2M r) \Phi_{,tt} - 2M r\Phi_{,tr} - \left( 2M r\Phi_{,t} \right)_{,r} - 2a\Phi_{,r\phi}
\nonumber\\ 
&& 
 - \left(\Delta\Phi_{,r} \right)_{,r} 
  - \frac{1}{\sin\theta}\left( \sin\theta\Phi_{,\theta} \right)_{,\theta}
  - \frac{1}{\sin^2\theta}\Phi_{,\phi\phi} = 0,
\qquad
\label{Eq:StationaryWaveEq}
\end{eqnarray}
where, here and in what follows, subindices following a coma refer to partial
derivatives; for instance \mbox{$\Phi_{,tr} = \partial_r\partial_t\Phi$}.

For a stationary solution (such that $h$ and $U^\mu$ are independent of $t$),
the scalar potential has the form
\begin{equation}
\Phi = e\left[ -t + \psi(r,\theta,\phi) \right],
\label{Eq:PhiAnsatz}
\end{equation}
with a new function $\psi$ which does not depend on $t$, and where the positive
constant $e$ corresponds to the Bernoulli constant (per unit mass), defined as
\begin{equation}
e = - h\, U_\mu K^\mu = - h\, U_t = -\Phi_{,t}, 
\end{equation}
where $K=\partial_t$ is the Killing vector field associated with the time
symmetry of Kerr spacetime.

Introducing the ansatz~\eqref{Eq:PhiAnsatz} into~\eq{Eq:StationaryWaveEq}
yields
\begin{equation}
\left(\Delta\psi_{,r} \right)_{,r} + \frac{1}{\sin\theta}\left( \sin\theta\,\psi_{,\theta} \right)_{,\theta}
  + \frac{1}{\sin^2\theta}\psi_{,\phi\phi} + 2\,a\,\psi_{,r\phi} = 2M.
\label{Eq:StationaryWaveEqReduced}
\end{equation}
Despite of the presence of the rotation parameter $a$, this equation can
be separated into radial and angular parts by means of a decomposition in
terms of the standard spherical harmonics $Y^{\ell m}(\theta,\phi)$:
\begin{equation}
\psi(r,\theta,\phi) = \sum\limits_{\ell m} R_{\ell m}(r) Y^{\ell m}(\theta,\phi),
\end{equation}
with the functions $R_{\ell m}$ to be determined. Introduced into
\eq{Eq:StationaryWaveEqReduced} this gives\footnote{For simplicity, we assume
that $Y^{00} = 1$ while for $\ell > 0$ the spherical harmonics $Y^{\ell m}$
are defined with the usual normalization.}
\begin{equation}
\frac{\ud}{\ud r} \left(\Delta\frac{\ud R_{00}}{\ud r} \right) = 2 M,
\label{Eq:StationaryWaveEql0}
\end{equation}
for $\ell = 0$, and
\begin{equation}
\frac{\ud}{\ud r} \left(\Delta\frac{\ud R_{\ell m}}{\ud r} \right) + 2\,i\,m\,a \frac{\ud R_{\ell m}}{\ud r} - \ell(\ell+1) R_{\ell m}
 = 0,
 \label{Eq:StationaryWaveEqRad}
\end{equation}
for $\ell\geq 1$. Integrating \eq{Eq:StationaryWaveEql0} once gives
$$
\frac{\ud R_{00}}{\ud r} = \frac{2M r + c_0}{(r- r_+)(r - r_-)},
$$
for some constant $c_0$, where $r_\pm = M \pm \sqrt{M^2 - a^2}$ denote the
roots of $\Delta$. In order for $R_{00}$ to be regular at the event horizon
$r = r_+$, one needs to choose $c_0 = -2M r_+$, such that the factor $r -
r_+$ in the denominator is canceled. This yields
\begin{equation}
R_{00} = 2M\ln\left( \frac{r - r_-}{r_+ - r_-} \right)
\label{Eq:R00}
\end{equation}
plus a constant which is irrelevant since the flow only depends on the
gradient of $\Phi$. Note that $R_{00}$ is regular for all $r > r_-$, but
diverges at the Cauchy horizon $r = r_-$.\footnote{Note also that  $R_{00}$
and its gradient diverge at the horizon in the extremal case $a = \pm M$,
when $r_+ = r_- = M$.} Therefore, the ``spherical" ($\ell = 0$) piece of
$\psi$ is fixed to the specific function~(\ref{Eq:R00}) by the requirement
of regularity at the horizon.

\eq{Eq:StationaryWaveEqRad} describes the ``non-spherical" ($\ell\geq 1$)
contributions to $\psi$ and can be brought into the hypergeometric differential
equation by introducing the dimensionless coordinate
\begin{equation}
x := \frac{r - r_+}{r_+ - r_-},
\end{equation}
which ranges from $-1$ to $\infty$ as $r$ varies from $r_-$ to
$\infty$ and is zero at the event horizon $r = r_+$. In terms of this,
\eq{Eq:StationaryWaveEqRad} reads
\begin{eqnarray}
&& x(1+x)\frac{\ud ^2 R_{\ell m}}{\ud x^2}
 + \left( 1 + 2x + \frac{2\,i\,m\,a}{r_+ - r_-} \right)\frac{\ud R_{\ell m}}{\ud x}
\nonumber\\ 
&& \qquad\qquad\qquad\qquad\qquad - \ell(\ell+1) R_{\ell m} = 0,
 \end{eqnarray}
which, after the further substitution $x = -y$, yields the standard form
of the hypergeometric differential equation (see, for example~\cite{DLMF},
Sec.~15). The solutions which are regular at the event horizon $x = 0$
can be written in terms of Gauss' hypergeometric function $F$ (as defined
in~\cite{DLMF}, Sec.~15):
\begin{equation}
R_{\ell m}(r) = A_{\ell m} F(-\ell,\ell+1; 1 + i\,m\,\alpha; -x),
\label{Eq:Rlm}
\end{equation}
where $A_{\ell m}$ is a free (complex) constant and where we have introduced
the dimensionless quantity
$$\alpha := \frac{2\,a}{r_+ - r_-} = \frac{a}{\sqrt{M^2 - a^2}}.$$ 
Note that \eq{Eq:Rlm} is actually a polynomial in $r$ of order
$\ell$,\footnote{These polynomials are related to the associated Legendre
functions of the first kind, see~\cite{lPsSsT88,DLMF}. For the special case
$c=1$ these polynomials are also related to the shifted Legendre polynomials.}
since for any complex number $c\neq 0,-1,-2,\ldots$,
\begin{equation}
F(-\ell,\ell+1;c;  -x) = \sum\limits_{n=0}^\ell \frac{(\ell + n)!}{(\ell - n)!} \frac{1}{(c)_n} \frac{x^n}{n!},
\end{equation}
with $(c)_n := c(c+1)(c+2)\cdots (c+n-1)$ for $n\geq 1$ and $(c)_0 := 1$. A
few examples relevant for this article are:
\begin{enumerate}
\item[$\ell=0$]: $F(0,1; c; -x) = 1$ (``spherical" Bondi-Michel-type accretion
which will be discussed in a future work)
\item[$\ell=1$]: $F(-1,2; c; -x) = 1 + \frac{2x}{c}$ (wind accretion discussed
in~\cite{lPsSsT88,vKrM93,tejeda18})
\item[$\ell=2$]: $F(-2,3; c; -x) = 1 + \frac{6x}{c} + \frac{12x^2}{c(c+1)}$
(choked accretion, discussed in the Schwarzschild limit in~\cite{TAH20},
and in the present paper for arbitrary rotation)
\end{enumerate}

Summarizing, the general solution describing a steady-state, irrotational
flow on a Kerr background which is regular at the horizon and which has an
ultrarelativistic stiff equation of state is characterized by the potential
\begin{equation}
\Phi = e \left[  -t + 2M\ln(1+x) + {\cal F}(r,\theta,\phi) \right],
\label{Eq:PSTSolution}
\end{equation}
with
\begin{equation}
{\cal F}(r,\theta,\phi) := \sum\limits_{\ell=1}^\infty\sum\limits_{m=-\ell}^\ell A_{\ell m}
F(-\ell,\ell+1; c;  -x)  Y^{\ell m}(\theta,\phi),
\label{Eq:PSTExpansion}
\end{equation}
where  we recall that $A_{\ell m}\in \Complex$, $x = (r - r_+)/(r_+ - r_-)$,
$c = 1 + i\,m\,\alpha$, and $\alpha = 2\,a/(r_+ - r_-)$.

Except for the addition of an irrelevant constant, the expression for the
potential in \eq{Eq:PSTSolution} agrees with Eq.~(30) in~\cite{lPsSsT88},
taking into account the relations~(\ref{Eq:tBL},\ref{Eq:phiBL}) between
the Kerr-type coordinates and the Boyer-Lindquist coordinates used in that
reference.

For $\Phi$ as given in \eq{Eq:PSTSolution} to be real, the coefficients
$A_{\ell m}$ need to satisfy the reality conditions
\begin{equation}
A_{\ell -m} = (-1)^m A_{\ell m}^*,
\label{Eq:RealityConditions}
\end{equation}
so that there are $2\ell + 1$ independent real constants for each $\ell$. Note
also that $F(-\ell,\ell+1; 1+i\,m\,\alpha;  0) = 1$ on the event horizon;
hence the coefficients $A_{\ell m}$ describe the $\ell m$ contributions of
the fluid potential $\Phi$ when evaluated on the horizon cross section.

The specific enthalpy and four-velocity are obtained from substituting
\eq{Eq:PSTSolution} into \eq{Eq:nU}, which yields
\begin{equation}
\begin{split}
\frac{h^2}{e^2} =\ & 1 + \frac{2M}{\varrho^2}\frac{r(r + r_+) + 2M r_+}{r - r_-} \\
 & + \frac{4M}{\varrho^2}\left( r_+{\cal F}_{,r} - \frac{a}{r - r_-}{\cal F}_{,\phi} \right) \\
&  - \frac{1}{\varrho^2}\left( \Delta\,{\cal F}_{,r}^2 + 2a\,{\cal F}_{,r}{\cal F}_{,\phi} + {\cal F}_{,\theta}^2
  + \frac{{\cal F}_{,\phi}^2}{\sin^2\theta} \right),
\end{split}
  \label{Eq:n2}
\end{equation}
and
\begin{subequations}
\begin{eqnarray}
\frac{h}{e} U^t &=& 1 + \frac{2M r}{\varrho^2}\frac{r + r_+}{r - r_-} + \frac{2M r}{\varrho^2} {\cal F}_{,r},
\label{Eq:Psit}\\
\frac{h}{e} U^r &=& \frac{1}{\varrho^2}\left( -2M r_+ + \Delta\,{\cal F}_{,r} + a\,{\cal F}_{,\phi} \right),
\label{Eq:Psir}\\
\frac{h}{e} U^\theta &=& \frac{1}{\varrho^2}{\cal F}_{,\theta},
\label{Eq:Psitheta}\\
\frac{h}{e} U^\phi &=& \frac{1}{\varrho^2}\left( \frac{2M a}{r - r_-} 
 + a\,{\cal F}_{,r} + \frac{1}{\sin^2\theta}{\cal F}_{,\phi} \right).
\label{Eq:Psiphi}
\end{eqnarray}
\end{subequations}

Recall that the gradient of $\Phi$ needs to be timelike for the solution to
be well-defined, which is equivalent to the requirement that the right-hand
side of \eq{Eq:n2} be positive. In general, this condition cannot be satisfied
everywhere outside the horizon. Since $F(-\ell,\ell+1; 1+i\,m\,\alpha;  -x)$
grows like $r^\ell$ at large distances, the right-hand side of \eq{Eq:n2} is
dominated by the term $-\Delta\,{\cal F}_{,r}^2/\varrho^2 \sim -r^{2\ell-2}$
for a solution containing multipoles up to a given $\ell$ and hence will
eventually become negative, for a sufficiently large radius, if $\ell\geq
2$. However, since $h^2/e^2 > 1$ when ${\cal F} = 0$, one can always choose
the coefficients $A_{\ell m}$ small enough such that the right-hand side
of \eq{Eq:n2} is positive (and hence $h$ well-defined) within a finite
spherical shell of the form $r_+\leq r \leq {\cal R}$ containing the horizon.

A further restriction comes from the requirement that the fluid should fall
into the black hole at the horizon, such that the four-velocity satisfies
the inequality
\begin{equation}
U^\mu\nabla_\mu r = U^r = \frac{e}{h}\frac{1}{\varrho^2}\left[ -2M r_+ + a\,{\cal F}_{,\phi} \right] < 0
\label{Eq:HorizonInflow}
\end{equation}
at the horizon $r = r_+$, which is equivalent to the bound $a\,{\cal
F}_{,\phi} < 2M r_+$ at $r = r_+$. We will show shortly that this is,
as expected, a consequence of the requirement for $\nabla^\mu\Phi$ to be
future-directed timelike.

\subsection{ZAMO frame and three-velocity}
\label{SubSec:ZAMO}

For the results and calculations that follow, it is convenient to express the
four-velocity in terms of an orthonormal frame instead of local coordinates. A
very convenient frame in the Kerr exterior spacetime is the one associated
with ZAMOs~\cite{jB70,jB73}, that is, observers whose world lines are tangent
to a linear combination of the Killing vector fields,
\begin{equation}
\frac{\partial}{\partial t} + \Omega\frac{\partial}{\partial\phi},
\quad \text{with} \quad \Omega = \frac{2M a r}{\Sigma},
\label{Eq:ZAMO}
\end{equation}
and
\begin{equation}
\Sigma = (r^2 + a^2)^2 - a^2\Delta\sin^2\theta = \Delta\varrho^2 + 2M r (r^2 + a^2) .
\end{equation}
The ZAMO's angular velocity $\Omega$ is singled out by the requirement of
zero angular momentum. These observers' tangent vectors are also orthogonal
to the $t_{\rm BL} = \co $ Boyer-Lindquist time slices, and in this sense
they generalize the ``local Eulerian observers'' used in~\cite{TAH20} to
discuss the quadrupolar flow in a Schwarzschild background.

A natural orthonormal frame associated with the ZAMOs is given by the following
basis vectors (see~\cite{jB70,jB73}):\footnote{Here and in the following,
hatted indices refer to labels for this orthonormal frame.}
\begin{subequations}
\begin{align}
e_{\hat t} &= \frac{1}{\varrho}\sqrt{\frac{\Sigma}{\Delta}}
\left( \frac{\partial}{\partial t} + \Omega\frac{\partial}{\partial\phi} \right),\\
e_{\hat r} &= \frac{\sqrt{\Delta}}{\varrho}\left( \frac{\partial}{\partial r} 
 + \frac{2M r}{\Delta} \frac{\partial}{\partial t} + \frac{a}{\Delta} \frac{\partial}{\partial \phi} \right),\\
e_{\hat \theta} &= \frac{1}{\varrho} \frac{\partial}{\partial \theta},\\
e_{\hat \phi} &= \frac{\varrho}{\sqrt{\Sigma}\sin\theta} \frac{\partial}{\partial \phi}.
\end{align}
\end{subequations}

The orthonormal components of the four-velocity are given by
\begin{subequations}
\begin{eqnarray}
\frac{h}{e} U^{\hat t} 
 &=& \frac{1}{\varrho}\sqrt{\frac{\Sigma}{\Delta}}\left( 1 - \Omega\, {\cal F}_{,\phi} \right),
\label{Eq:hu0}\\
\frac{h}{e} U^{\hat r}
  &=& \frac{1}{\varrho\sqrt{\Delta}}\left( -2M r_+ + \Delta\,{\cal F}_{,r} + a\,{\cal F}_{,\phi} \right),
\label{Eq:hu1}\\
\frac{h}{e} U^{\hat \theta} &=& \frac{1}{\varrho} {\cal F}_{,\theta},
\label{Eq:hu2}\\
\frac{h}{e} U^{\hat \phi} &=& \frac{\varrho}{\sqrt{\Sigma}\sin\theta} {\cal F}_{,\phi}.
\label{Eq:hu3}
\end{eqnarray}
\end{subequations}
On the other hand, the components of the three-velocity are defined as
\begin{subequations}
\begin{eqnarray}
V^{\hat r} &=& \frac{U^{\hat r}}{U^{\hat t}}  = \frac{-2M r_+ + \Delta\,{\cal F}_{,r} + a\,{\cal F}_{,\phi}}{\sqrt{\Sigma}\left( 1 - \Omega\, {\cal F}_{,\phi} \right)},
\label{Eq:hV1}\\
V^{\hat \theta} &=& \frac{U^{\hat \theta}}{U^{\hat t}}  = \sqrt{\frac{\Delta}{\Sigma}}\, \frac{ {\cal F}_{,\theta}}{ 1 - \Omega\, {\cal F}_{,\phi} },
\label{Eq:hV2}\\
V^{\hat \phi} &=& \frac{U^{\hat \phi}}{U^{\hat t}}  =  \frac{\varrho^2\sqrt{\Delta}}{\Sigma\,\sin\theta} \,\frac{ {\cal F}_{,\phi}}{ 1 - \Omega\, {\cal F}_{,\phi} } ,
\label{Eq:hV3}
\end{eqnarray}
\end{subequations}
with the corresponding Lorentz factor
\begin{equation}
 \Gamma = U^{\hat t} =  \frac{1}{\sqrt{1-V^2}},
\end{equation}
where
\begin{equation}
 V = \sqrt{(V^{\hat r})^2+(V^{\hat \theta})^2+(V^{\hat \phi})^2}.
\end{equation}

A number of interesting conclusions can be drawn from these representations
of the four- and three-velocities. First, the four-velocity vector is
future-directed timelike outside the horizon if and only if $U^{\hat t} >
0$ and if the magnitude of the three-velocity $V$ is smaller than one. This
is equivalent to the two conditions
\begin{equation}
\Omega\, {\cal F}_{,\phi} < 1
\label{Eq:FutureDirected}
\end{equation}
and
\begin{equation}
\begin{split}
V^2 = \frac{1}{\Sigma( 1 - \Omega\, {\cal F}_{,\phi} )^2} & \Bigg[
\left( -2M r_+ + \Delta\,{\cal F}_{,r} + a\,{\cal F}_{,\phi} \right)^2 \\
&  + \Delta\left( {\cal F}_{,\theta}^2 + \frac{\varrho^4}{\Sigma\sin^2\theta}{\cal F}_{,\phi}^2 
 \right) \Bigg] < 1.
\end{split}
 \label{Eq:Timelike}
\end{equation}
In the axisymmetric case, when ${\cal F}_{,\phi} = 0$, the first inequality
is automatically satisfied and the second one simplifies considerably:
\begin{equation}
V^2 = \frac{(2M r_+ - \Delta\,{\cal F}_{,r})^2 + \Delta\,{\cal F}_{,\theta}^2}{\Sigma} < 1.
\label{Eq:TimelikeAxiSym}
\end{equation}
Since $\Sigma \geq  (2Mr_+)^2 + \Delta\varrho^2$ for $r\geq r_+$,
one can always satisfy this inequality for small enough values
of the gradient of ${\cal F}$. The restrictions implied by the
inequalities~(\ref{Eq:FutureDirected},\ref{Eq:Timelike}) for a quadrupolar
solution ($\ell = 2$) will be analyzed in more detail in the next two sections.

The next property that can be inferred from
Eqs.\,(\ref{Eq:FutureDirected},\ref{Eq:Timelike}) is obtained by taking the
limit $r\to r_+$. In this limit, the inequality~(\ref{Eq:FutureDirected})
yields $\Omega_+ \left. {\cal F}_{,\phi} \right|_{r=r_+}\leq 1$, where
$\Omega_+ = a/(2M r_+)$ is the angular velocity of the event horizon. This
provides a bound for the value of ${\cal F}_{,\phi}$ at the horizon, and
comparison with \eq{Eq:HorizonInflow} reveals the meaning of this bound:
the fluid cannot flow out of the black hole, a property that is, of course,
expected on physical grounds!
By requiring that the four-velocity $U^\mu$ is everywhere timelike on the
horizon, one can further eliminate the possibility that $\Omega_+ {\cal
F}_{,\phi} = 1$ somewhere on the horizon; otherwise \eq{Eq:HorizonInflow}
would imply that $U^\mu$ is tangent to the horizon and thus cannot be
timelike. Summarizing, the requirement for $U^\mu$ to be future-directed
timelike at the horizon yields the strict inequality,
\begin{equation}
\Omega_+ \left. {\cal F}_{,\phi} \right|_{r=r_+} < 1
\end{equation}
which implies that the flow can only cross inwards the event horizon.

Another point to notice from the expressions for the three-velocity of the fluid in
\eqs{Eq:hV1}-\eqref{Eq:hV3} is that the fluid is at rest with respect to a
ZAMO if and only if the function ${\cal F}$ satisfies
\begin{equation}
\Delta\,{\cal F}_{,r} +a\,{\cal F}_{,\phi} = 2M r_+,\qquad
{\cal F}_{,\theta} = {\cal F}_{,\phi} = 0.
\label{Eq:ZAMORest}
\end{equation}

Finally, we note that, even though the ZAMO frame is very useful in many
situations, this frame is not well-defined at the event horizon nor in
the region inside the black hole between the two horizons $r_-$ and $r_+$,
where $\Delta \leq 0$. In case one is interested in analyzing the flow at
or inside the horizon, one may use instead the orthonormal frame adapted to
local Eulerian observers relative to the $t = \co $ Kerr-type coordinates.

\subsection{Conserved quantities}
\label{SubSec:Conservation}

Due to the presence of the Killing vector fields $K = \partial_t$ and $L
= \partial_\phi$ of the Kerr spacetime, the following four-currents are
divergence-free:
\begin{subequations}
\begin{eqnarray}
J^\mu = \rho\, U^\mu,\\
J_{\cal E}^\mu = -T^\mu{}_\nu K^\nu ,\\
J_{\cal L}^\mu = T^\mu{}_\nu L^\nu ,
\end{eqnarray}
\end{subequations}
corresponding to the rest-mass, energy, and angular momentum current
densities, respectively.

For an ultrarelativistic stiff fluid, the specific enthalpy $h = 2K \rho$
is proportional to the particle density and \mbox{$P = \rho\,h/2$}, such that
\begin{subequations}
\begin{eqnarray}
J^\mu &=&  \rho\, U^\mu = \frac{\rho}{h}\nabla^\mu\Phi, \\
T^\mu{}_\nu &=& \rho\, h\, U^\mu U_\nu + P\,\delta^\mu{}_\nu \nonumber \\ 
&=& \frac{\rho}{h}\left[
 (\nabla^\mu\Phi)(\nabla_\nu\Phi) - \frac{1}{2}\delta^\mu{}_\nu(\nabla^\alpha\Phi)(\nabla_\alpha\Phi) \right]. 
 \nonumber \\
\end{eqnarray}
\end{subequations}
In particular, using Eqs.~(\ref{Eq:PSTSolution},\ref{Eq:Psir})
we find
\begin{subequations}
\begin{eqnarray}
J^r &=& \frac{\rho\, e}{h}\frac{1}{\varrho^2}\left[
 -2M r_+ + \Delta\,{\cal F}_{,r} + a\,{\cal F}_{,\phi} \right],
\label{Eq:Jr}\\
J_{\cal E}^r &=& e\, J^r,\\
J_{\cal L}^r &=& e\, {\cal F}_{,\phi} J^r.
\end{eqnarray}
\end{subequations}

Since the flow is stationary, the equation  $\nabla_\mu J^\mu = 0$ gives
\begin{equation}
( \varrho^2\sin\theta J^r)_{,r} + ( \varrho^2\sin\theta J^\theta)_{,\theta}
 + ( \varrho^2\sin\theta J^\phi)_{,\phi} = 0.
\end{equation}
Therefore, the mass accretion rate (current flux) associated with $J$
through a two-surface $S$ is given by
\begin{equation}
\dot{M} = -\int\limits_S (J^r N_r + J^\theta N_\theta + J^\phi N_\phi) \varrho^2\sin\theta \,\ud S,
\label{Eq:Flux}
\end{equation}
with $(N_r,N_\theta,N_\phi)$ the unit outward normal field and $\ud S$
a differential area element of $S$. If $S$ is closed, then $\dot{M}$ is
independent of any deformations of $S$, since $J^\mu$ is conserved. For
example, if $S$ is a constant-$r$ surface, then
\begin{equation}
\dot{M} = -\int\limits_S J^r\varrho^2\sin\theta \,\ud\theta \,\ud\phi,
\label{Eq:Flux2}
\end{equation}
which is independent of $r$.
Using now the orthogonality relations of the spherical harmonics we can
integrate \eq{Eq:Flux2} as
\begin{equation}
\dot{M} =  8\pi M r_+\frac{\rho \,e}{h}
 = 4\pi(r_+^2 + a^2)\frac{\rho\, e}{h},
\label{Eq:ParticleAccretionRate}
\end{equation}
which is constant since $\rho/h = 1/(2K)$.

Similarly, for the energy accretion rate we have
\begin{equation}
\dot{\cal E} = -\int\limits_S J_{\cal E}^r \,\varrho^2\sin\theta\, \ud\theta\, \ud\phi 
 = 4\pi(r_+^2 + a^2)\frac{ \rho\, e^2}{h}
 = e\dot{M},
\label{Eq:EnergyAccretionRate}
\end{equation}
while, for the angular momentum accretion rate
\begin{eqnarray}
\dot{J} &=& -\int\limits_S J_{\cal L}^r \,\varrho^2\sin\theta \ud\theta \ud\phi 
\nonumber\\
 &=& -a\frac{\rho\, e^2}{h}\int\limits_{r = r_+} {\cal F}_{,\phi}^2 \sin\theta \ud\theta \ud\phi
 \nonumber\\
 &=& -a\frac{\rho\, e^2}{h}\sum\limits_{\ell=1}^\infty \sum\limits_{m =
 -\ell}^\ell m^2 | A_{\ell m} |^2.
\label{Eq:AngularMomentumAccretionRate}
\end{eqnarray}

Notice that the mass and energy accretion rates are uniquely determined by
the $\ell = 0$ part of the solution (they are independent of the coefficients
$A_{\ell m}$), which in turn was determined by the regularity requirement at
the event horizon. In contrast to this, the angular momentum accretion rate
is solely determined by the $\ell > 0$ part of the solution. Interestingly,
the sign of $\dot{J}$ indicates that the accreted material always slows
down the spin of the black hole (unless the flow is perfectly axisymmetric
in which case $\dot{J} = 0$). Therefore, the accretion flow described
by~(\ref{Eq:PSTSolution}) always drives the Kerr black hole away from
extremality ($|J|$ decreases, $M$ increases, such that $J/M^2$ decreases).

\section{The axisymmetric quadrupolar flow}
\label{Sec:AxisymmetricQuadFlow}

In this section we shall focus on the axisymmetric quadrupolar solution,
i.e.~the velocity potential $\Phi$ in \eq{Eq:PSTSolution} for which all
of the coefficients  $A_{\ell m}$ vanish except for the $(\ell,m) = (2,0)$
contribution, which results in
\begin{equation}
\Phi = e\left[ -t + 2M\ln\left( \frac{r - r_-}{r_+ - r_-}\right) + A\, F(r,\theta,\phi) \right],
\label{s3e1}
\end{equation}
with
\begin{equation}
F(r,\theta,\phi) = \left(3\,r^2 - 6Mr + 2M^2 + a^2 \right)(3\cos^2\theta - 1),
\label{s3e2}
\end{equation}
where, as we shall see below, $e$ can be identified as a scaling factor for
the gas' thermodynamic state while $A$ determines the overall flow morphology.

We can now exploit all of the results derived in the previous section. In
particular, from \eqs{Eq:hV1}--\eqref{Eq:hV3}, we obtain the following
expressions for the spatial components of the three-velocity as described
by the ZAMOs
\begin{subequations}
\begin{eqnarray}
V^{\hat r} &=&  \frac{-2Mr_+ + A\Delta F_{,r} }{\sqrt{\Sigma}} , 
\label{s3e3} \\
V^{\hat \theta} &=& \sqrt{\frac{\Delta}{\Sigma}} A \,F_{,\theta},
 \label{s3e4} \\
V^{\hat \phi} &=& 0,
\label{s3e5} 
\end{eqnarray}
\end{subequations}
where
\begin{subequations}
\begin{eqnarray}
F_{,r} &=& 6(r-M)(3\cos^2\theta - 1) ,
\label{Eq:Fr}\\
F_{,\theta} &=& - 6\left(3\,r^2 - 6Mr + 2M^2 + a^2 \right)\cos\theta\sin\theta .\nonumber \\
& & 
\label{Eq:Ftheta}
\end{eqnarray}
\end{subequations}

The value for the constant $e$ can be set by specifying a reference point at
which the fluid state is known. Calling $h_0$  the specific enthalpy and $V_0$
the magnitude of the three-velocity at this reference point, from \eq{Eq:hu0},
we have
\begin{equation}
e = \Gamma_0 h_0  \varrho_0 \sqrt{\frac{\Delta_0 }{\Sigma_0}},
\label{s3e6}
\end{equation}
where $\Gamma_0 = 1/\sqrt{1-V_0^2}$. 

Using Eqs.~(\ref{Eq:n2}) and (\ref{Eq:hu0}) we find that the specific enthalpy
in this case is given by 
\begin{eqnarray}
\frac{h^2}{e^2} &=& 
 \frac{\Sigma(1-V^2)}{\Delta\varrho^2}
\nonumber\\
 &=&  1 + \frac{2Mr}{\varrho^2} + \frac{4M^2}{\varrho^2}\, \frac{r+r_+}{r-r_-}
\nonumber\\
  && \quad + \frac{4Mr_+}{\varrho^2} A \,F_{,r}
 - \frac{A^2}{\varrho^2}\left( \Delta F_{,r}^2  + F_{,\theta}^2\right).
\label{Eqh}
\end{eqnarray}

On the other hand, denoting by $\rho_0$ the rest-mass density at the reference
point, from the equation of state we have $\rho/\rho_0 = h/h_0$. Using this,
and substituting \eq{s3e6} back into \eq{Eq:hu0}, we obtain
\begin{equation}
 \frac{\rho}{\rho_0} = \frac{h}{h_0} =  \frac{\Gamma_0 \varrho_0}{\Gamma \varrho}
 \sqrt{\frac{\Sigma}{\Sigma_0}\frac{\Delta_0}{\Delta}}.
\label{s3e7}
\end{equation}
From \eq{s3e7}, we note that the following combination of variables
\begin{equation}
 \rho\,\Gamma\varrho \sqrt{\frac{\Delta}{\Sigma}} = \co
\end{equation}
yields a global constant that characterizes the resulting flow. Indeed,
as follows from \eqs{Eq:ParticleAccretionRate} and (\ref{Eq:hu0}), this constant is proportional to the total mass
accretion rate. 
Also note that, from \eq{Eqh}, it is clear that both $h$
and $\rho$ are completely regular (finite) quantities at the event horizon
($r=r_+$),\footnote{Provided that $|A|$ remains sufficiently small. See
the discussion below~\eq{Eq:Ellipsoid} for conditions on $A$ that guarantee
that $h^2/e^2 > 0$ near the horizon.} although they do become infinite at
the Cauchy horizon ($r=r_-$).

Provided that $A\neq0$, the flow structure described by \eqs{s3e3}-\eqref{s3e5}
consists of an inflow-outflow morphology. We can characterize this morphology
in terms of the location of the stagnation points, i.e.~points at which
the three-velocity vanishes. From Eqs.~(\ref{s3e4},\ref{Eq:Ftheta})
we see that $V^{\hat \theta}$
vanishes only at points along the polar axis ($\theta = 0,\,\pi$) and on the
equatorial plane ($\theta = \pi/2$). Now it only remains examining the points
at which $V^{\hat r}=0$ restricted to either $\theta = 0,\,\pi$ or $\theta =
\pi/2$. From Eqs.~(\ref{s3e3},\ref{Eq:Fr}) we can distinguish two qualitatively different cases:

\emph{Case 1:} When $A>0$, the resulting structure consists of inflow across
an equatorial region and outflow confined to the polar regions (bipolar
outflow). In this case, $V^{\hat r}$ vanishes at two points
along the polar axis symmetrically located with respect to the origin at a
coordinate distance $r=\cal S$ that satisfies
\begin{equation}
 A = \frac{Mr_+}{6({\cal S}-r_-)({\cal S}-r_+)({\cal S}-M)}.
 \label{eA1}
\end{equation}

\begin{figure}
 \begin{center}
 \includegraphics[width=\linewidth]{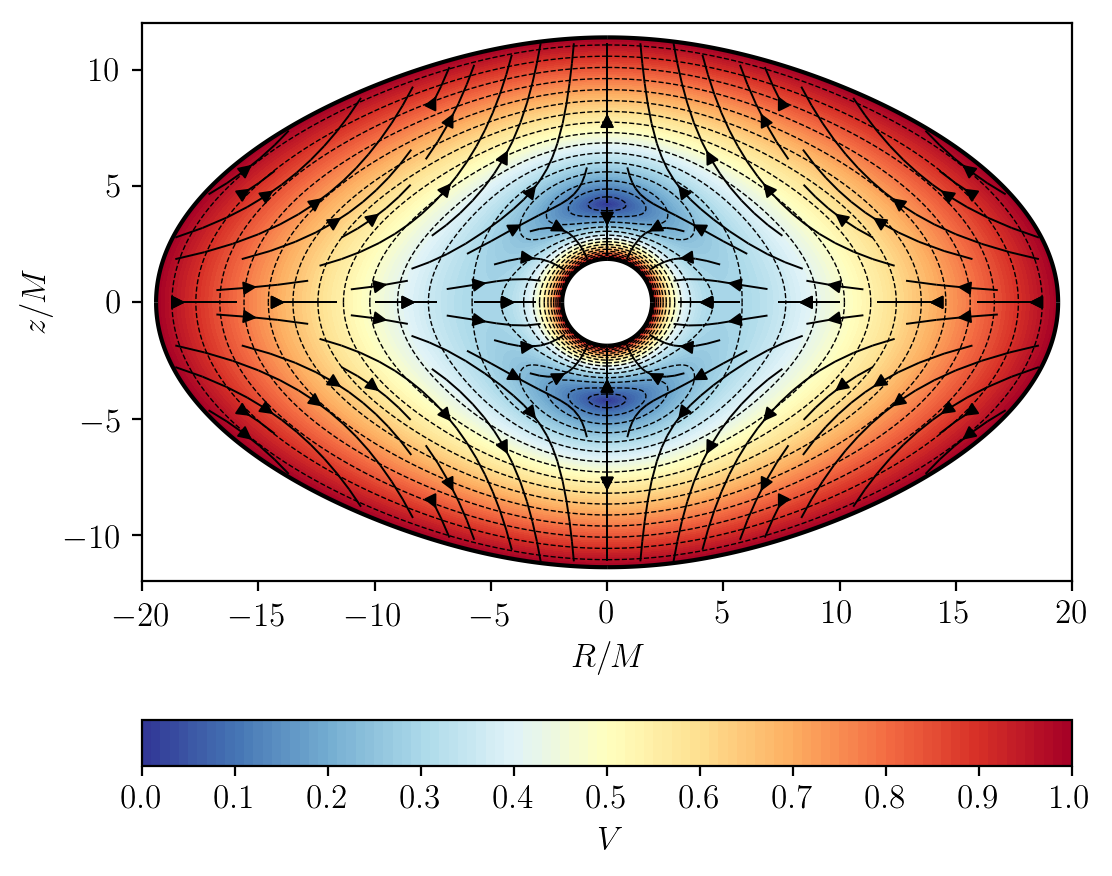}
 \end{center}
 \caption{Example of the axisymmetric quadrupolar flow with $AM=0.01$ and a
 central Kerr black hole with $a=0.5$. The stagnation points in this case are
 located along the polar axis at a coordinate distance $r={\cal S} \simeq
 4.2 M$. The figure shows isocontours of the three-velocity's magnitude
 $V$. Note that $V$ becomes luminal at the event horizon ($r=r_+$) and at
 the outer ellipsoid indicated by a black, thick line. Fluid streamlines
 are indicated by thick, solid lines with an arrow. The axes correspond to
 the  cylindrical-like coordinates $R = \sqrt{r^2+a^2}\,\sin\theta$, $z =
 r\,\cos\theta$. }
\label{f1}
\end{figure}

See Fig.~\ref{f1} for an example of the resulting flow for  $AM=0.01$
(which corresponds to ${\cal S} \simeq 4.2 4 M$) and a Kerr black hole with
$a = 0.5M$.

\emph{Case 2:} When $A<0$, the scenario is reversed and one has two bipolar
inflow regions and outflow across the equatorial region.  In this case, we
have that $V^{\hat r}$ vanishes now at an infinite number of points located
on an equatorial ring of radius  $r = \cal S$ satisfying
\begin{equation}
 A = -\frac{Mr_+}{3({\cal S}-r_-)({\cal S}-r_+)({\cal S}-M)}.
\label{eA2}
 \end{equation}

\begin{figure}
 \begin{center}
 \includegraphics[width=\linewidth]{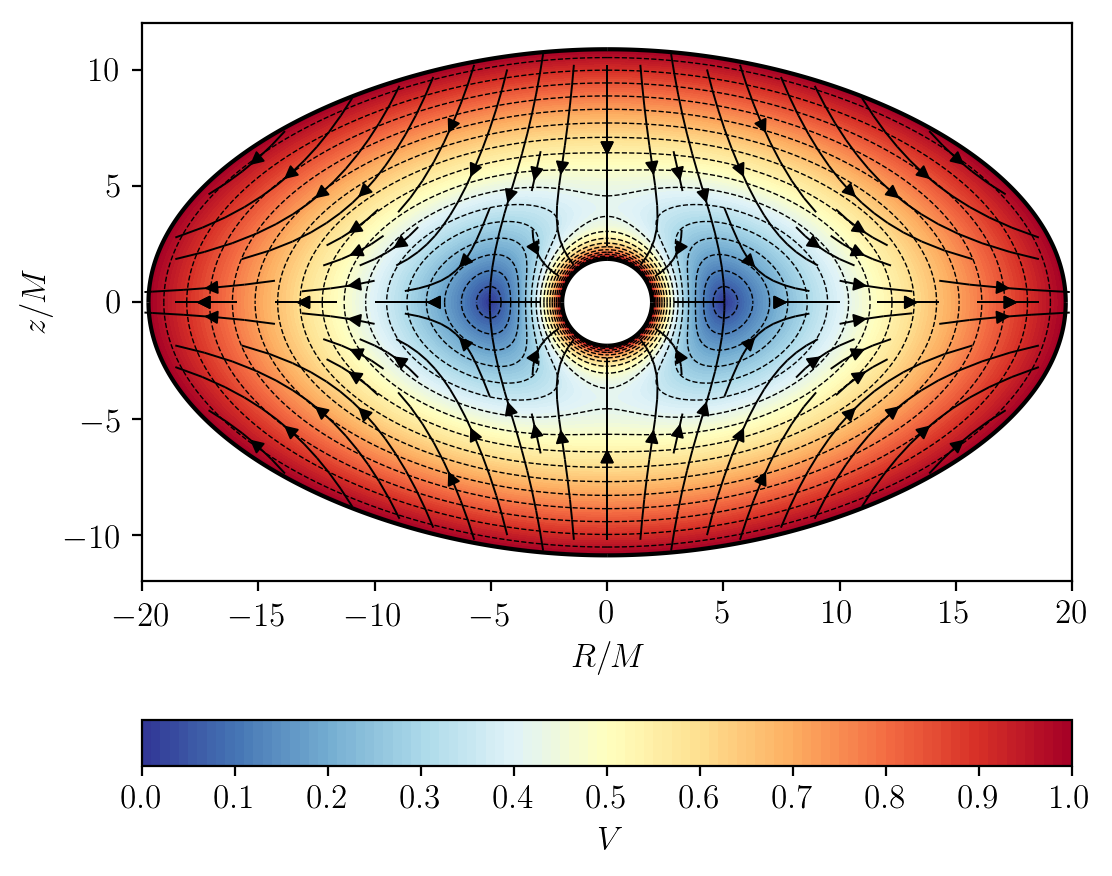}
 \end{center}
 \caption{Same as in Fig.~\ref{f1}, except that $AM = -0.01$ is negative. The
 stagnation points in this case are located on an equatorial ring at a
 coordinate distance $r = {\cal S} \simeq 5 M$. 
 }
\label{f2}
\end{figure}

In Fig.~\ref{f2}, we show an example of the resulting flow for  $AM = -0.01$
(which corresponds to ${\cal S} \simeq 5 M$) and a Kerr black hole with $a
= 0.5M$.

In both examples shown in Figs.~\ref{f1} and ~\ref{f2}, it is apparent that
$V$ becomes luminal at two surfaces. From \eqs{s3e3}-\eqref{s3e5}, and as
discussed in the previous section, it is simple to see that one such surface is
the black hole's event horizon located at $r=r_+$. This behavior is, however,
a coordinate effect related to the fact that the ZAMOs become ill defined at
this radius. Indeed, using \eqs{Eq:Psit}-\eqref{Eq:Psiphi}, it can be seen
that the fluid's four-velocity is completely regular across the event horizon.

On the other hand, the outer surface at which $V=1$ signals an unavoidable
characteristic of the quadrupolar solution.
This surface, that in what follows we shall refer to as $\cal E$, marks the
transition of the gradient $\Phi_{,\mu}$ from being timelike (for points inner
to $\cal E$) to becoming spacelike (for points outside $\cal E$). Moreover,
from \eq{s3e7} we see that, at this surface, the density $\rho$ becomes zero
and, for points outside $\cal E$, $\rho$ ceases to be a real quantity.
For these reasons, we have to consider $\cal E$ as the outermost boundary
delimiting the spatial domain of applicability of the quadrupolar solution.

An expression for $\cal E$ can be obtained by combining \eqs{Eqh} and
\eqref{s3e7}, and rewrite the condition $V^2 = 1$ as the following second
order polynomial in $\cos^2\theta$:
\begin{equation}
 c_2(r)\,\cos^4\theta + c_1(r)\,\cos^2\theta + c_0(r) = 0,
 \label{e_poly}
\end{equation}
where
\begin{subequations}
\begin{eqnarray}
c_0(r) &=& r^2 + 2Mr + 4M^2\left(\frac{r+r_+}{r-r_-}\right) 
\nonumber \\
& & 
 -24 A M r_+ (r-M) - 36A^2\Delta(r-M)^2,\\
c_1(r) &=& a^2 + 72A(r - M)M r_+  
\nonumber \\
& & - 36 A^2 \big[\left(3\,r^2 - 6Mr + 2M^2 + a^2 \right)^2 
\nonumber \\
& & - 6 (r-M)^2 \Delta \big]  , \\
c_2(r) &=& 36 A^2 \left(M^2-a^2\right) \left(3\,r^2 - 6Mr + 4M^2 - a^2 \right).\nonumber \\
& & 
\end{eqnarray}
\end{subequations}

From \eq{e_poly}, one can show that, in the limit \mbox{$AM\ll 1$} (which
necessarily implies ${\cal S} \gg M$ and $r \gg M$), $\cal E$ reduces to
the simple ellipsoid of revolution described by
\begin{equation}
 r^2\left(1 + 3\,\cos^2\theta\right) = x^2+y^2+ 4\,z^2 = \frac{1}{(6A)^2},
\label{Eq:Ellipsoid}
\end{equation}
where, within this same limit, from \eq{eA1}  in Case 1  we have $A =
M^2/(3\,{\cal S}^3)$ while, from \eq{eA2} in Case 2 it follows that $A =
-2M^2/(3\,{\cal S}^3)$.

\begin{figure}
 \begin{center}
 \includegraphics[width=\linewidth]{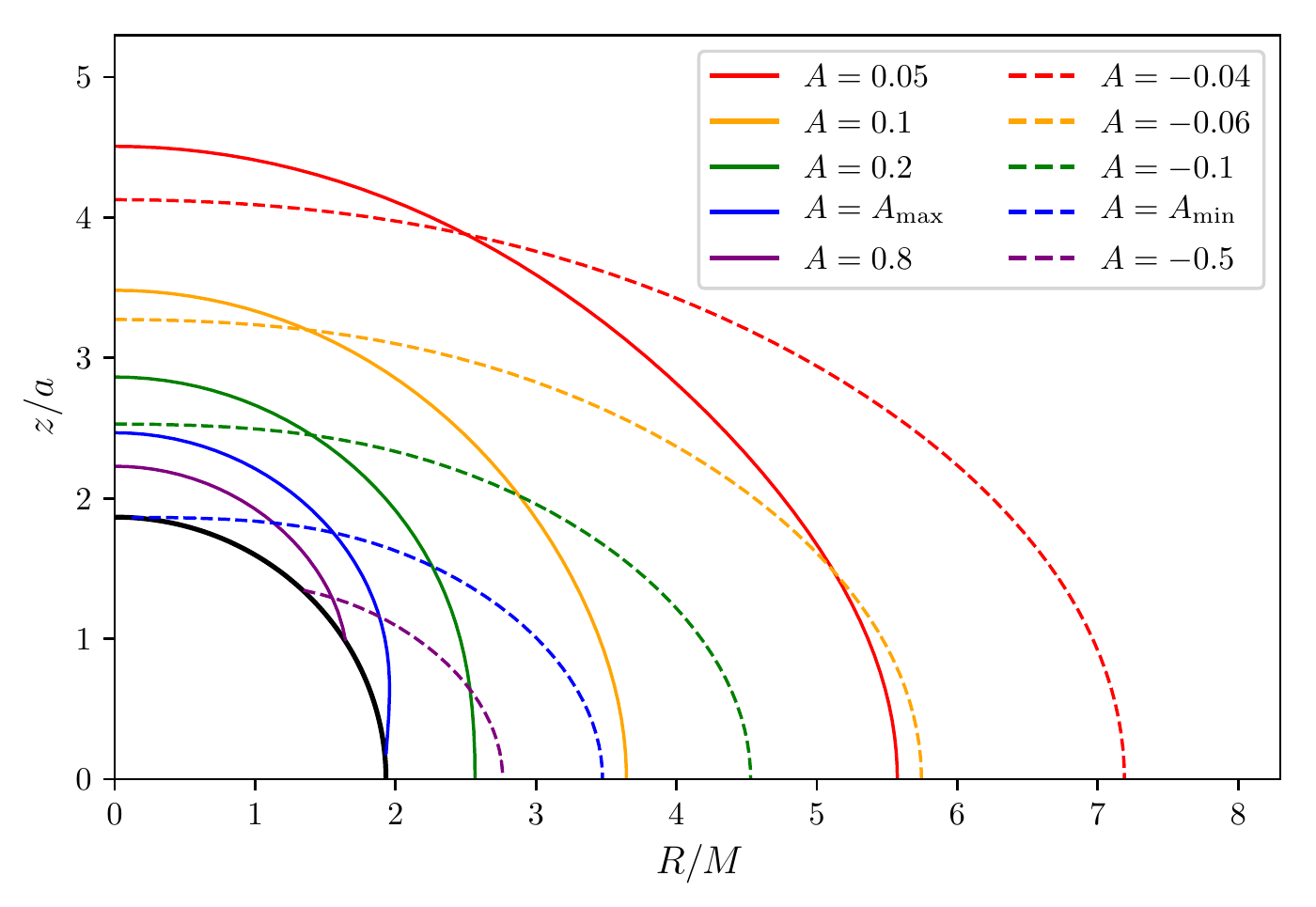}
 \end{center}
 \caption{ Outermost boundary $\cal E$ of the axisymmetric, quadrupolar flow
 for several values of the coefficient $A$. Solid, colored lines represent
 cases with $A>0$, while dashed colored lines correspond to $A<0$. The central
 black hole has a spin parameter $a = 0.5M$. The event horizon is indicated
 by a solid black line. Note that the curves $\cal E$ corresponding to  $A
 = A_{\max} \simeq 0.41$ and  $A = A_{\min} \simeq -0.21$ touch the event
 horizon at the equator and pole, respectively. Curves with $A>A_{\max}$
 or $A<A_{\min}$  actually pierce through the horizon. The axes correspond
 to the cylindrical-like coordinates $R = \sqrt{r^2+a^2}\,\sin\theta$, $z =
 r\,\cos\theta$. }
\label{f3}
\end{figure}%

On the other hand, by examining \eq{e_poly}, it becomes apparent that, for
a sufficiently large value of $|A|$, the surface $\cal E$ actually pierces
through the event horizon. When $A>0$, $\cal E$ first touches the horizon at
$\theta=\pi/2$ while, when $A<0$, $\cal E$ starts merging with the horizon
at $\theta=0$. This means that  the coefficient $|A|$ cannot be arbitrarily
large or, in other words, that there is a minimum possible value ${\cal
S}_{\min}$ for  ${\cal S}$ such that ${\cal S}_{\min} > r_+$. In order to
find the maximum value $A_{\max}$, let us first substitute $\theta=\pi/2$
in \eq{e_poly} and then evaluate the result at $r=r_+$. Doing this gives
the condition $c_0(r_+) = 0$, which can be solved explicitly for $A$ as
\begin{equation}
A_{\max} = \frac{5M^2 - a^2 + 3M\sqrt{M^2-a^2}}{24M(M^2-a^2)}.
\label{eAmax}
\end{equation}
Similarly, for finding the minimum value $A_{\min}$, we substitute $\theta=0$
in \eq{e_poly} and then evaluate the result at $r=r_+$. This results in the
condition $c_2(r_+) +  c_1(r_+) + c_0(r_+) = 0$ which can be solved for $A$ as
\begin{equation}
A_{\min} = - \frac{M^2 + M \sqrt{M^2-a^2}}{12M(M^2-a^2)}.
\label{eAmin}
\end{equation}

For a Schwarzschild black hole, $A_{\max} = 1/3$ (${\cal S} \simeq 2.32M$)
and $A_{\min} = -1/6$ (${\cal S} \simeq 2.80M$). On the other hand, for a
Kerr black hole with $a=0.5M$, we have $A_{\max} \simeq 0.41$ (${\cal S}
\simeq 2.18M$) and $A_{\min} = -0.21$ (${\cal S} \simeq 2.61M$). Finally,
note that in the extremal limit \mbox{$a\rightarrow M$}, $A$ actually becomes
unbounded, i.e~$(A_{\min},\, A_{\max})\rightarrow(-\infty,\,\infty)$. In
Fig.~\ref{f3} we show examples of the boundary $\cal E$ for different values
of $A$ for a Kerr black hole with $a = 0.5M$.

\begin{figure}
 \centering
 \includegraphics[width=\linewidth]{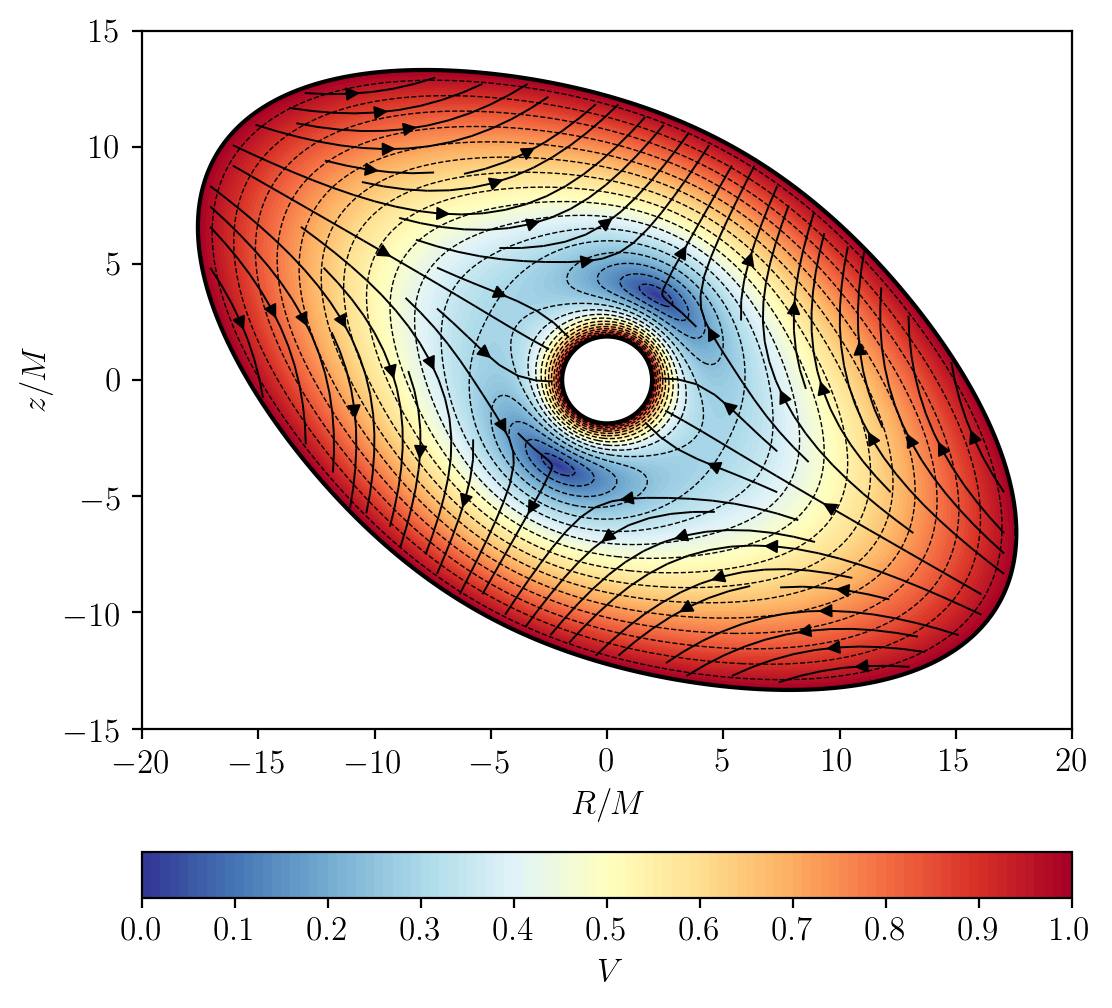}
 \caption{Same as in Fig.~\ref{f1}, except that now we consider an inclination
 angle $\theta_0 = 30^\circ$. In order to show the stagnation points,
 the plot corresponds to the plane $\phi = \phi(\epsilon) = -0.15882$
 (see Table~\ref{tab:stagnation} and accompanying discussion). }
 \label{fig:misaligned}
\end{figure}

We conclude this section with some words regarding the case in which there is
a misalignment between the accretion flow morphology and the black hole spin
axis. As we show in further detail in the Appendix~\ref{App:Misaligned}, this
case still allows for the same kind of inflow-outflow solutions. However, the
resulting expressions become more involved as lack of axisymmetry forces us
to consider, in addition to the $(\ell,m) = (2,0)$ mode, the contributions
from the $m=-2,-1,1,2$ modes. As an example of the resulting accretion
flow, in Fig.~\ref{fig:misaligned} we show the result of considering a
misalignment angle of $\theta_0 = 30^\circ$ for the same flow parameters as
in Fig.~\ref{f1}.

\section{Choked accretion}
\label{Sec:choked}

Here we apply the results obtained in the previous section to
the choked accretion scenario discussed in~\cite{TAH20} for a Schwarzschild
spacetime. The idea is the following: a gas flow is injected radially inwards
from points lying close to the equator of a sphere of certain coordinate radius
$r = {\cal R} > r_+$ (the ``injection sphere") toward the black hole. Part
of this flow will be accreted by the black hole and disappears through the
event horizon. However, when the injection rate is sufficiently large, it
has been shown in~\cite{TAH20} that (due to an anisotropic density field)
part of the flow is diverted and ejected toward the poles.
Under these conditions, the resulting flow is characterized by an inflow
region originating from an equatorial belt in the injection sphere and a
bipolar outflow region (Case 1 discussed in the previous section).

For the reasons mentioned in the introduction, we shall limit the rest of this
work to the case in which the black hole's angular momentum is perpendicular
to the injection plane, that is, the equator of the injection sphere lies
inside the equatorial plane $\theta = \pi/2$ of the Kerr spacetime.

\begin{figure}[t]
 \centering
 \includegraphics[width=0.45\textwidth]{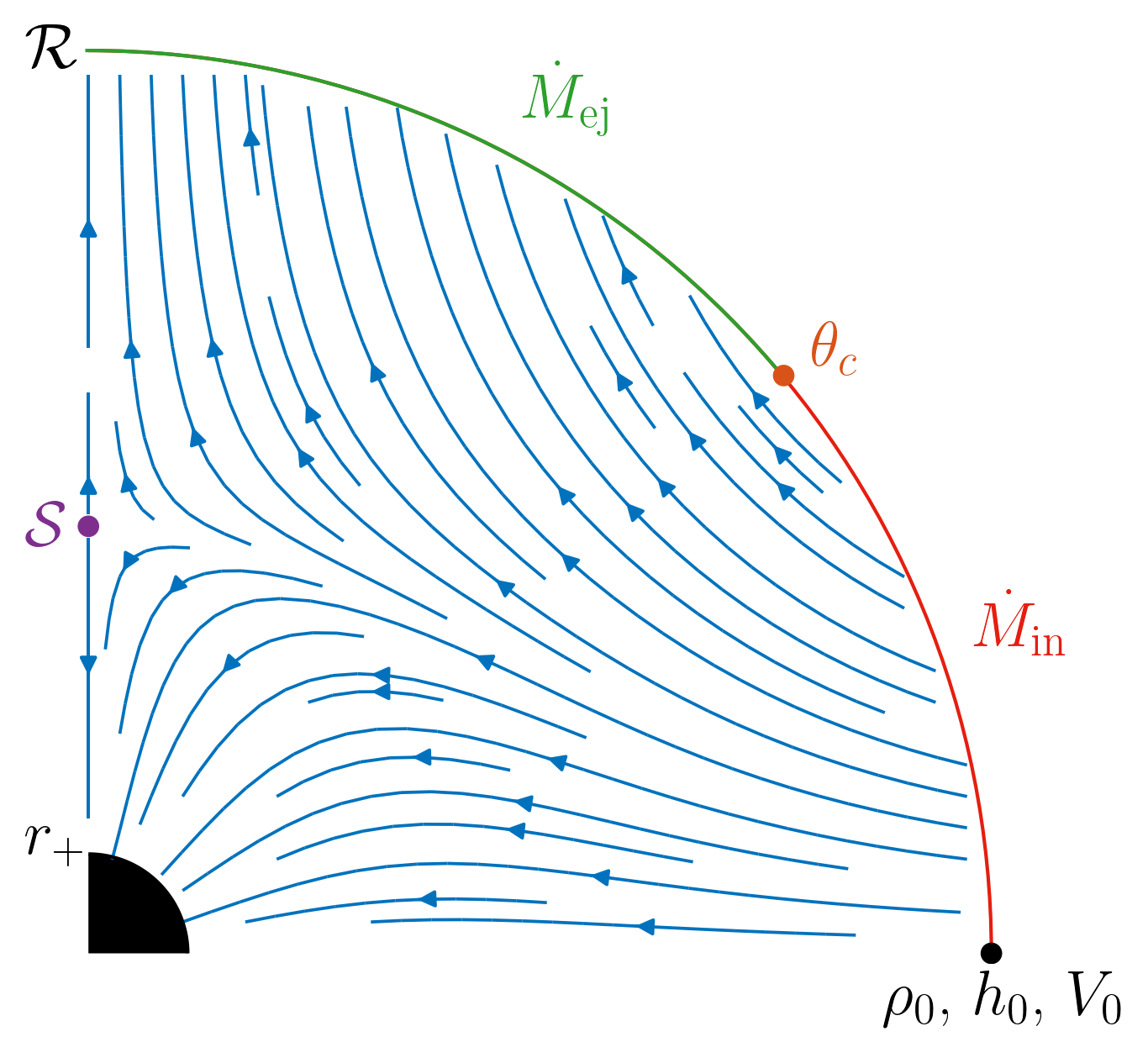}
 \caption{Schematic representation of the choked accretion model in the
 polar plane. Shown are the black hole region ($r < r_+$), the location of
 the injection sphere ($r = \mathcal{R})$, the location of the reference
 point $(r= \mathcal{R},\,\theta=\pi/2)$ where the data $(\rho_0,h_0,V_0)$
 characterizing the solution are specified, the location of the stagnation
 point at $r = \mathcal{S}$ and the critical angle $\theta_c$ which separates
 the inflow from the outflow regions on the injection sphere.}
 \label{fig:setup}
\end{figure}

For given values of the black hole parameters $(M,a)$, we characterize the
resulting flow by specifying the fluid properties at the equator of the
injection sphere, i.e., at $r=\mathcal{R},\ \theta = \pi/2$.
At this reference point, we prescribe the thermodynamic variables $\rho_0 =
\rho(\mathcal{R},\pi/2)$, $h_0 = h(\mathcal{R},\pi/2)$, and the magnitude
of the fluid's three-velocity $V_0$ as measured by a ZAMO at this location. See
Fig.~\ref{fig:setup} for a schematic representation of the setup.

By imposing these boundary conditions in \eqs{s3e3} and \eqref{s3e6}, it
follows that
\begin{subequations}
\begin{eqnarray}
 e &=& \Gamma_0 h_0  \mathcal{R} \sqrt{\frac{\Delta_0 }{\Sigma_0}},
 \label{Eq:e} \\
 A &=& \frac{\sqrt{\Sigma_0}\,V_0 - 2Mr_+ }{6(\mathcal{R}-M)\Delta_0},
 \label{Eq:A1}
\end{eqnarray}
\end{subequations}
where
\begin{subequations}
\begin{eqnarray}
\Delta_0 &=& (\mathcal{R}-r_-)(\mathcal{R}-r_+),\\
\Sigma_0 &=& (\mathcal{R}^2 + a^2)\mathcal{R}^2 + 2M\mathcal{R}a^2, \\
\Gamma_0 &=& (1-V_0^2)^{-1/2}.
\end{eqnarray}
\end{subequations}

With the values for the model parameters in \eqs{Eq:e} and \eqref{Eq:A1}, all
of the results derived in the previous section can be directly adopted. In
particular, the velocity field of the corresponding solution is given by
\eqs{s3e3}-\eqref{s3e5}, the fluid enthalpy by \eq{Eqh}, and the density
field by \eq{s3e7}. Also note that the location of the stagnation points in
this case follows by combining \eq{eA1} and \eq{Eq:A1}, which results in
\begin{equation}
   \frac{\sqrt{\Sigma_0}\,V_0}{Mr_+} - 2  = \frac{({\cal R}-r_-)({\cal R}-r_+)({\cal R}-M)}{({\cal S}-r_-)({\cal S}-r_+)({\cal S}-M)}.
\end{equation}
This equation can be explicitly solved for $\cal S$ as
\begin{equation}
\begin{split}
 \mathcal{S} = M  + & \left( \xi + \sqrt{\xi^2-\frac{(M^2-a^2)^3}{27}} 
\right)^{1/3} \\
+ & \left( \xi - \sqrt{\xi^2-\frac{(M^2-a^2)^3}{27}} \right)^{1/3} ,
\end{split}
\end{equation}
where
\begin{equation}
 \xi = 
\frac{({\cal R}-r_-)({\cal R}-r_+)({\cal R}-M)Mr_+}{2(\sqrt{\Sigma_0}\,V_0 - 2 Mr_+)}.
\end{equation}

Finally note that, following a procedure analogous to that described in
\cite{TAH20}, one can obtain an expression for the projection of a streamline
onto the $r$-$\theta$ plane given by
\begin{equation}
\Psi = \cos\theta\left[1 + \frac{(r-r_-)(r-r_+)(r-M)}{(\mathcal{S}-r_-)(\mathcal{S}-r_+)(\mathcal{S}-M)}\frac{\sin^2\theta}{2} \right],
\end{equation}
where $\Psi$ is an integration constant. Streamlines with $|\Psi|<1$ accrete
onto the central black hole, those with $|\Psi|>1$ escape along the bipolar
outflow, while those with $\Psi=1$ ($\Psi=-1$) are connected to the stagnation
point at $\theta = 0$ ($\theta = \pi$).

\subsection{Parameter range}
\label{Sec:pspace}

The solution described by \eq{s3e1} with $e$ and $A$ as given in
\eqs{Eq:e}-\eqref{Eq:A1} is characterized by six parameters: $M$ and $a$
describing the black hole, and $\cal R$, $\rho_0$, $h_0$ and $V_0$ specifying
the boundary conditions at the injection sphere. As discussed in \cite{TAH20},
the obtained solution is actually scale-free with respect to the model
parameters $M$ (that sets the overall length scale),  $\rho_0$, and $h_0$
(that set the thermodynamic state of the fluid).

Once a Kerr background metric has been fixed with $M$ and $a$ (satisfying
$|a|<M$), our next goal is to determine the range for the parameters $\cal R$
and $V_0$ leading to solutions that:
\begin{enumerate}
 \item Are well-defined within the domain $r\in [r_+,\ {\cal R}]$.
 \item Present the inflow-outflow morphology of the choked accretion mechanism. 
\end{enumerate}
To this end, it is convenient to examine the ejection velocity defined as
\begin{equation}
 V_{\rm ej} \equiv V^{\hat r}({\cal R},\,0) = \frac{2V_0\sqrt{\Sigma_0} - 6Mr_+}{{\cal R}^2 + a^2},
\label{Eq:Vej}
\end{equation}
where we have used \eqs{s3e3} and \eqref{Eq:A1}.

Condition 1 is satisfied by requiring that the gradient of the potential
function remains timelike within the domain of interest, which is equivalent to
the condition that the right-hand side of \eq{Eqh} is positive for all $r\in
[r_+,{\cal R}]$ and all $\theta\in [0,\pi]$. In Appendix~\ref{App:Bounds}
we prove that this can be guaranteed by requiring
\begin{equation}
{\cal R}\geq 3M + r_+
\end{equation}
and demanding that $V_{\rm ej} < 1$. From \eq{Eq:Vej}, this last condition
in turn is equivalent to
\begin{equation}
 V_0 < \frac{{\cal R}^2 + a^2 + 6Mr_+}{2\sqrt{\Sigma_0}}.
\end{equation}

On the other hand, since we have already assumed inflow across the equator
of the injection sphere, condition~2 is satisfied by requiring $V_{\rm ej}
> 0$. Again, from \eq{Eq:Vej}, this condition translates as
\begin{equation}
 V_0 >  \frac{3Mr_+}{\sqrt{\Sigma_0}}.
\end{equation}
Therefore, the injection velocity parameter is restricted as
\begin{equation}
V_0 \in (V_{\min},\, V_{\max}), 
\end{equation}
with
\begin{equation}
 V_{\min} = \frac{3Mr_+}{\sqrt{\Sigma_0}}, \quad  V_{\max} = \frac{{\cal R}^2 + a^2}{2\sqrt{\Sigma_0}} + V_{\min}.
 \label{bounds}
\end{equation}

\subsection{Mass accretion, injection, and ejection rates}
\label{sec:MinMej}

\begin{figure*}
 \begin{center}
 \includegraphics[width=0.49\linewidth]{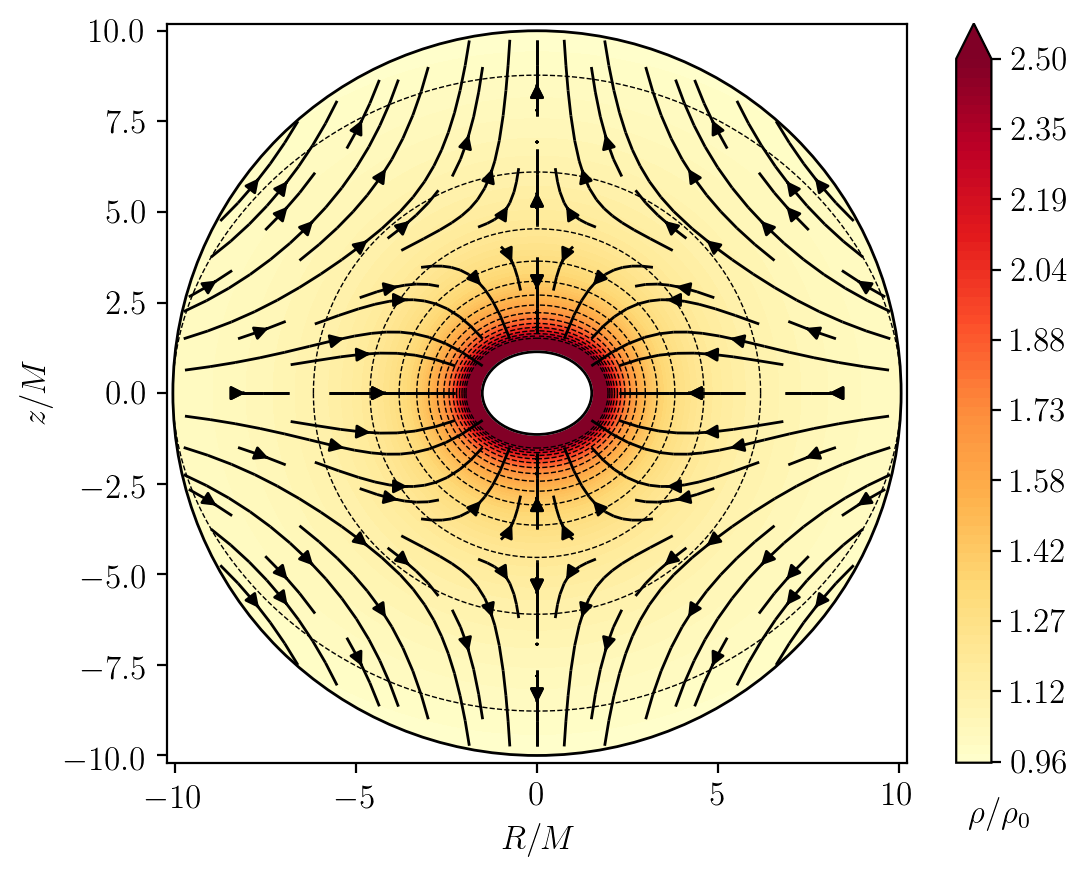}
 \includegraphics[width=0.49\linewidth]{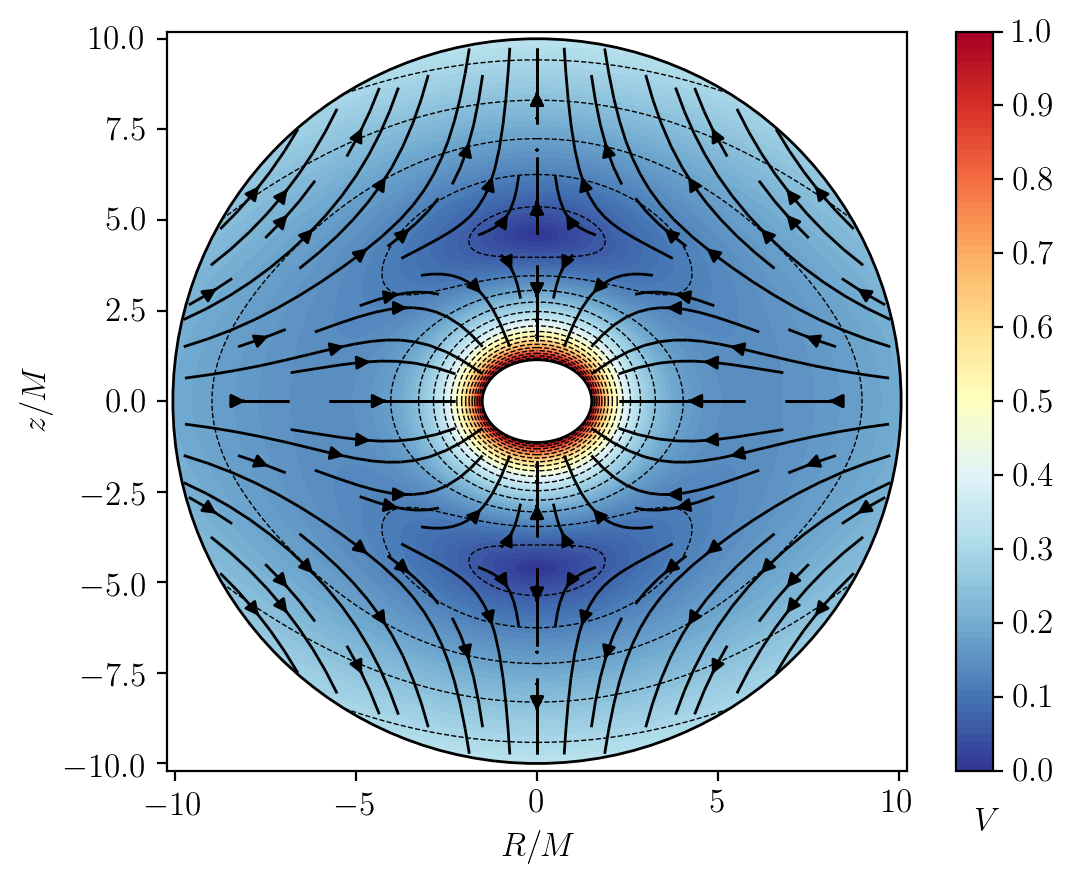}  
 \end{center}
 \caption{Analytic model of choked accretion for a Kerr black hole with
 $a=0.99\, M$ and flow parameters $\mathcal{R}= 10\, M$ and $V_0 = 0.2$.
 The figure shows isocontours of the fluid's normalized rest-mass density
 (left panel) as well as the magnitude of the three-velocity (right panel). The
 stagnation points are located on the symmetry axis with radius $\mathcal{S}
 \simeq 4.6011\, M$. Fluid streamlines are indicated by thick, solid lines
 with an arrow. The axes correspond to the cylindrical-like coordinates $R =
 \sqrt{r^2+a^2}\,\sin\theta$ and $z = r\,\cos\theta$.}
 \label{fstiff}
\end{figure*}

The accretion rate follows by substituting \eq{Eq:e} into
\eq{Eq:ParticleAccretionRate}, which results in
\begin{equation}
 \dot{M} =  8\pi M r_+\frac{\rho\,e}{h}
 = 8\pi M r_+ \Gamma_0\, \rho_0\,\mathcal{R} \sqrt{\frac{\Delta_0 }{\Sigma_0}}.
\end{equation}

By considering the flux of mass across the injection sphere, we can distinguish
between the inflow and outflow fluxes, $\dot{M}_{\rm in}$ and $\dot{M}_{\rm
ej}$, respectively, defined in such a way that
\begin{equation}
 \dot{M}_{\rm in} - \dot{M}_{\rm ej} = \dot{M}.
 \label{e5.2}
\end{equation}

We can calculate both fluxes explicitly by examining the radial component
of the fluid velocity at the injection sphere. From \eq{s3e3}, it follows
the existence of a critical angle $\theta_c$ given by
\begin{equation}
\theta_c 
 = \arccos\left[ 3 \left(1 - \frac{ 2Mr_+}{\sqrt{\Sigma_0} V_0} \right) \right]^{-1/2},
\end{equation}
such that there is inflow ($V^{\hat r} < 0$) for the equatorial belt defined
by $ \theta \in (\theta_c,\,\pi -\theta_c)$ and outflow  ($V^{\hat r} >
0$) for  the polar regions $\theta \in (0,\,\theta_c)$ and $\theta \in (
\pi -\theta_c ,\,\pi)$.

We can thus calculate $\dot{M}_{\rm in}$ in terms of $\theta_c$ as
\begin{equation}
\dot{M}_{\rm in} = -4\pi\int_{\theta_c}^{\pi/2} \rho\, U^r \varrho^2\sin\theta \,\ud\theta 
=  \Lambda\dot{M} 
\end{equation}
where
\begin{equation}
 \Lambda = \frac{2\cos^3\theta_c}{3\cos^2\theta_c - 1}
 = \frac{\sqrt{\Sigma_0} V_0}{ 3\sqrt{3} Mr_+}\left( 1 - \frac{ 2Mr_+}{\sqrt{\Sigma_0} V_0} \right)^{-1/2}.
\end{equation}
Clearly, in view of \eq{e5.2}, it follows that
\begin{equation}
\dot{M}_{\rm ej} = (\Lambda - 1) \dot{M} .
\end{equation}

In Fig.~\ref{fstiff} we show the isocontour levels of the rest-mass
density field, as well as the magnitude of the three-velocity $V$, and
the resulting fluid streamlines (black solid arrows)  for a
representative case with model parameters $a=0.99\, M$, $\mathcal{R}
= 10\, M$, and $V_0 = 0.2$.

\begin{figure}
 \begin{center}
 \includegraphics[width=0.42\textwidth]{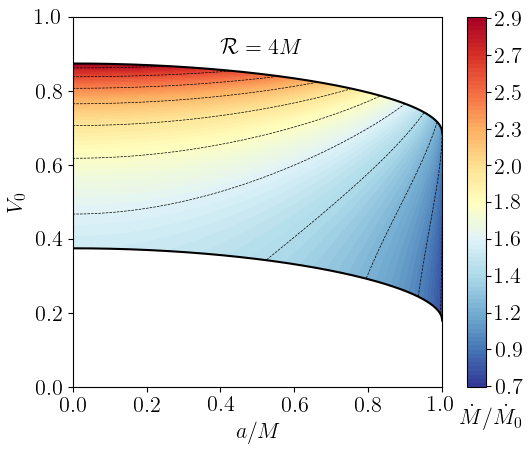}\\
 \includegraphics[width=0.42\textwidth]{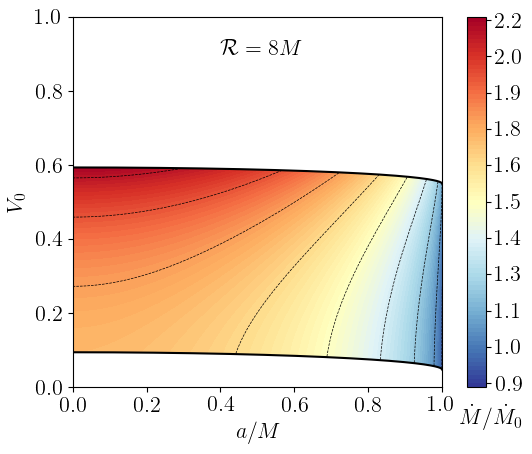}\\
 \includegraphics[width=0.42\textwidth]{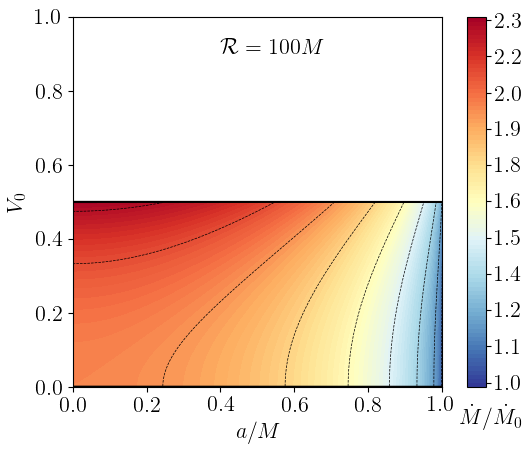}  
 \end{center}
 \caption{Mass accretion rate as a function of the model parameters
 ($a$,\,$V_0$) in units of $\dot{M}_0 = 8\pi M^2 \rho_0$. The value of the
 injection radius ${\cal R}$ in each case is indicated by a central label
 on each panel. The solid, black lines in each panel indicate the range of
 validity of the model parameters according to $V_0\in(V_{\min},\,V_{\max})$,
 with the lower boundary corresponding to $V_{\min}$ and the upper one
 to $V_{\max}$.}
 \label{fdotM}
\end{figure}

In Fig.~\ref{fdotM} we represent the regions in the parameter space
$(a,\,V_0)$ that lead to the choked accretion solution as discussed
in Sec.~\ref{Sec:pspace}.  The plotted isocontours correspond to the
mass accretion rate $\dot{M}$ expressed in units of $\dot{M}_0 = 8\pi M^2
\rho_0$.  Each panel corresponds to a different value of the injection radius
${\cal R}$, from top to bottom ${\cal R}/M = 4,\,8,\,100$.  The boundary
lines delimiting each region correspond to the $V_{\min}$ and  $V_{\max}$
limits given in \eq{bounds}.  From this figure we can see a general trend for
increasing values of $\dot{M}$ as the value of $V_0$ increases, while $\dot{M}$
decreases as the spin parameter $a/M$ grows from zero to 1. Also note that
the dependence on $a$ of the limits $V_{\min}$ and  $V_{\max}$ becomes less
noticeable as increasingly larger values of ${\cal R}$ are considered.

\begin{figure}
\centering
\includegraphics[width=0.48\textwidth]{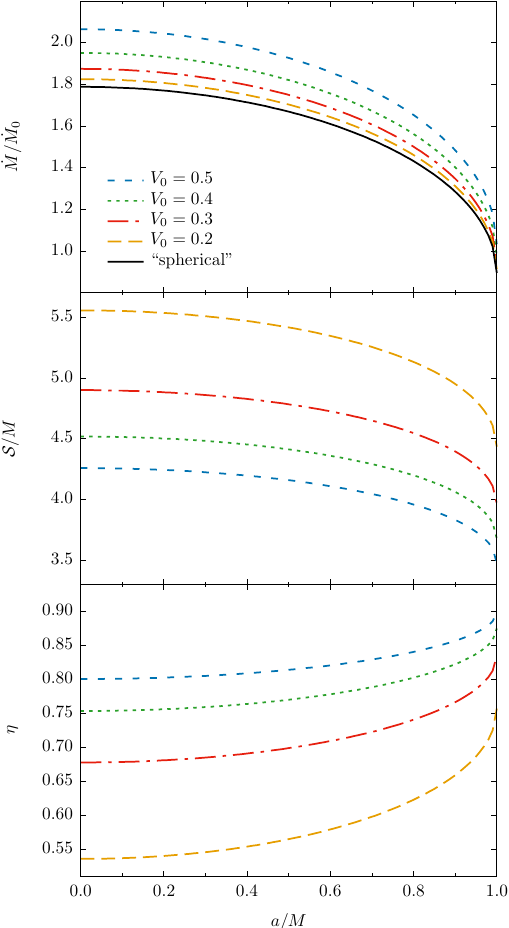}
\caption{Dependence of different properties of the choked accretion model
on the spin parameter $a/M$ and the injection velocity $V_0$, for an injection
sphere at $\mathcal{R} = 10\, M$. From top to bottom, each panel shows: the
mass accretion rate $\dot{M}$ in units of $\dot{M}_0 = 8\pi M^2\rho_0$, the
location of the stagnation points $\mathcal{S}/M$, and the ejection-to-injection
mass rate ratio $\eta=\dot{M}_\mathrm{ej}/\dot{M}_\mathrm{in}$. The black line in the
first panel corresponds to the ``spherical'' case $(\ell,m) = (0,0)$, for which $V_0 = 2Mr_+/\sqrt{\Sigma_0}\, \simeq 0.0003.$}
\label{fig:analytic}
\end{figure}

In Fig.~\ref{fig:analytic} we show three different properties of the
choked accretion model as a function of the spin parameter $a/M$, for
an injection sphere at $\mathcal{R} = 10\, M$. Each color line represents
a different value of the injection velocity. The quantities correspond
to: the mass accretion rate $\dot{M}$ (in units of $\dot{M}_0$) in the
top panel, the location of the stagnation points $\mathcal{S}$ in the
middle panel, and the ejection-to-injection mass rate ratio $\eta =
\dot{M}_{\rm{ej}}/\dot{M}_{\rm{in}}$ in the bottom panel. For comparison,
in the top panel we also show, in a black solid line, the accretion rate for
the ``spherically symmetric'' case $(\ell,m)=(0,0)$ corresponding to $A = 0$
and $V_0 = 2Mr_+/\sqrt{\Sigma_0}$ (note that there is no ejection for $V_0$
in the range between this value and $V_{\min}$). 

From the previous discussion we note that, as the injection velocity grows
from $V_{\min}$ to $V_{\max}$, we have:
\begin{itemize}
 \item The radii of the stagnation points decrease from ${\cal S} = {\cal R}$ to ${\cal
 S}_{\min}$.
 \item The critical angle increases from $\theta_c = 0$ to 
\begin{equation}
\theta_{\max} = \arccos\left[3\left(\frac{{\cal R}^2 + a^2 + 2Mr_+ }{{\cal R}^2 + a^2 + 6Mr_+}\right)\right]^{-1/2},
\end{equation}
 that, in the limit ${\cal R}\gg M$, converges to $\theta_{\max} = \arccos(1/\sqrt{3}) \simeq 54.7^{\circ}$.
 \item The mass injection rate increases from $\dot{M}_{\rm in}= \dot{M}$ to 
\begin{equation}
\dot{M}_{\rm in} = \frac{\left[ \frac{1}{3}\left( 1 + \frac{4Mr_+}{({\cal R} - r_+)({\cal R} - r_-)} \right) \right]^{3/2}}{\frac{2Mr_+}{({\cal R} - r_+)({\cal R} - r_-)}} \dot{M}.
\end{equation}
\end{itemize}

On the other hand, from Figs.~\ref{fdotM} and \ref{fig:analytic}, we note
that, as the spin parameter $a/M$ increases from zero to 1, the mass accretion
rate onto the central black hole decreases down to $\sim 50\%$, the location
of the stagnation point $S$ decreases by a factor of $\sim 10\%$, while the
ejection-to-injection mass rate ratio $\eta$ increases by up to $\sim 30\%$.

The analytic model studied in the previous sections allows us to explore
in detail the effect of the black holes's rotation on the choked accretion
mechanism. Unfortunately, this model cannot easily be extended to perform a
more general study including a more realistic equation of state. Keeping the
irrotational assumption one can still formulate the problem in terms of
a scalar potential; however, this potential satisfies a wave equation which
is nonlinear for a realistic equation of state. Clearly, this makes it much
harder to find an analytic treatment. For this reason, in the next section,
we extend our study to the case of a general polytropic fluid by performing
numerical simulations of the choked accretion scenario.

\section{Numerical simulations}
\label{Sec:Numeric}

\begin{figure*}
 \centering
 \includegraphics[width=0.49\textwidth]{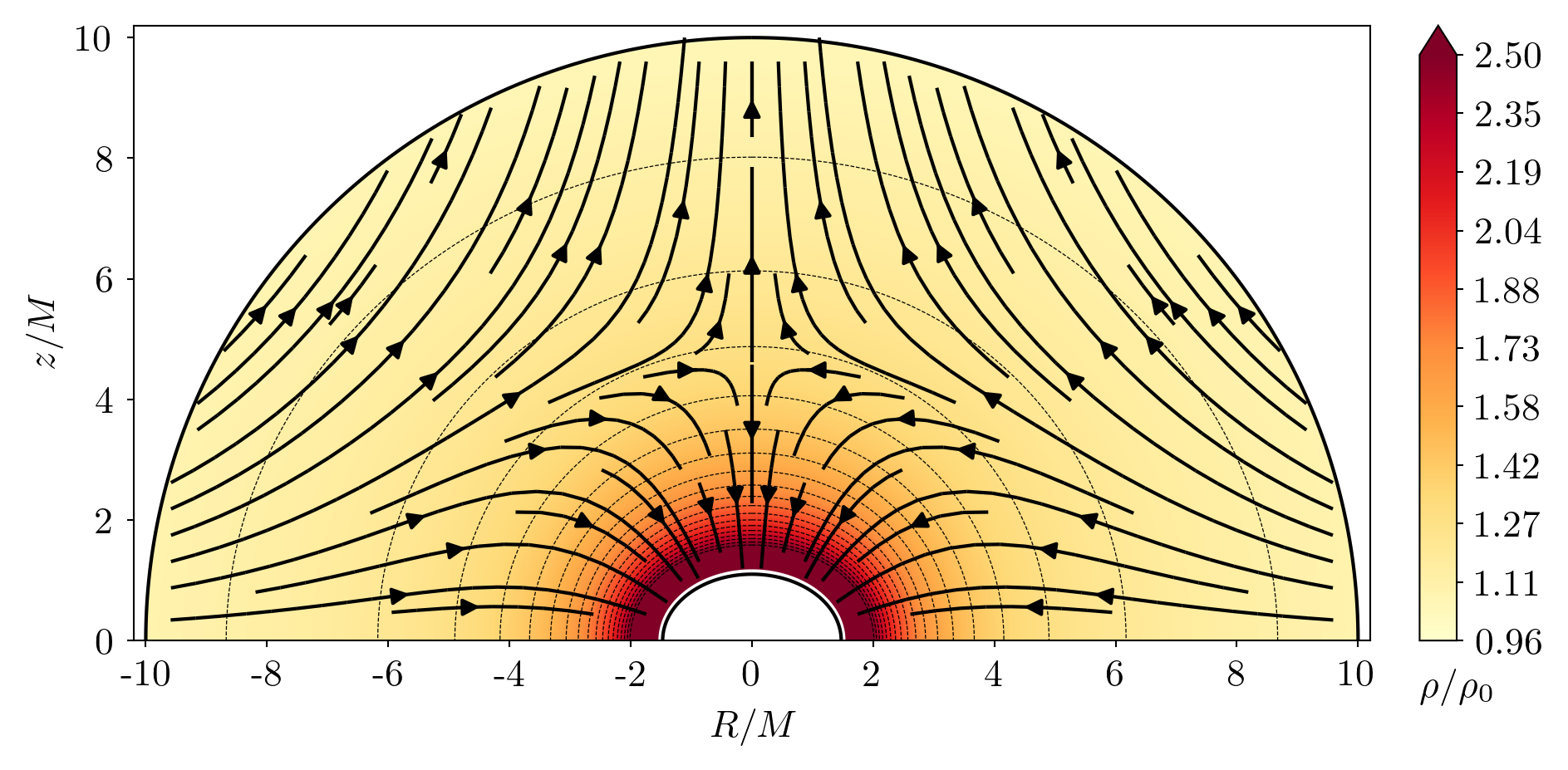}
 \includegraphics[width=0.49\textwidth]{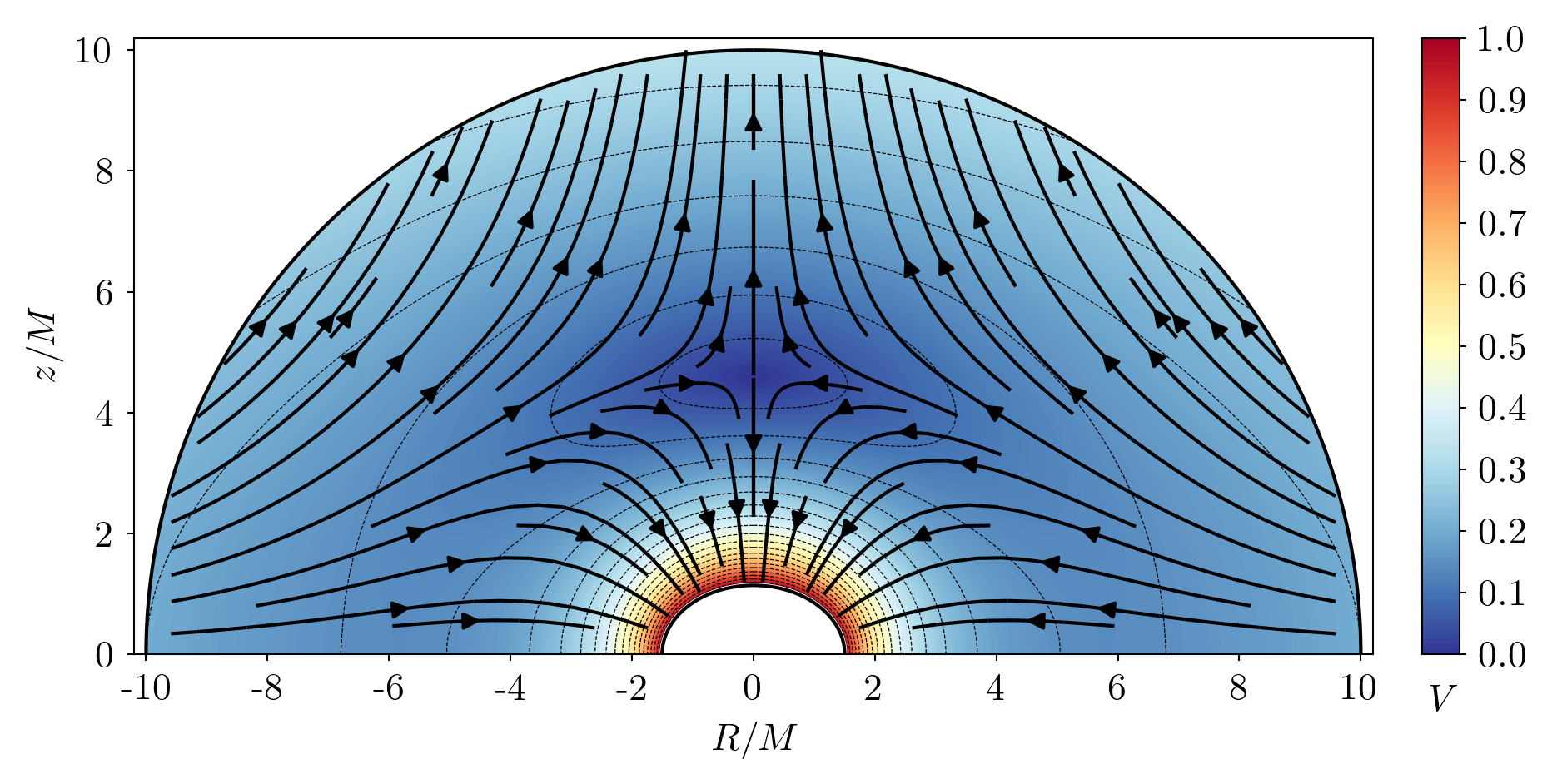}
 \caption{
 Validation test of the \textsc{aztekas} code. In this figure we show
 the steady state of numerical simulation for the benchmark test, which
 corresponds to the analytic solution presented in Sec.~\ref{Sec:choked},
 with parameters $\mathcal{R} = 10\, M$, $V_0 = 0.2$, and $a=0.99M$
 (compare with Fig~\ref{fstiff}). The figure shows the isocontour levels
 of the normalized rest-mass density $\rho/\rho_0$ (left panel) and the
 magnitude of the three-velocity (right panel) $V$, as measured by a ZAMO at
 this location. The fluid streamlines are indicated with black solid arrows.
 The simulation reached the stationary state at $t=180\, M$, showing a good
 agreement with the analytic solution.
 }
 \label{fig:comp-analytic}
\end{figure*}

The general solution presented by Petrich, Shapiro and
Teukolsky~\cite{lPsSsT88} revisited in Sec.~\ref{Sec:SteadyStateKerr} and, in
particular, the choked accretion scenario discussed in Sec.~\ref{Sec:choked},
are limited by the assumption of an ultrarelativistic gas with a stiff
equation of state, which leads to an unphysical speed of sound. In this
section we show, by means of full hydrodynamic numerical simulations,
that the main features of the choked accretion model are maintained when
the adopted equation of state is extended to consider a general polytropic
gas. Moreover, we make use of the analytical solution presented above as a
2D benchmark test for the validation of the code.

We perform full hydrodynamic numerical simulations with the
open source code \textsc{aztekas}\footnote{The code can be
downloaded from \url{https://github.com/aztekas-code/aztekas-main}.}
\citep{aguayo18,eTaA19}, which solves the general relativistic
hydrodynamic equations using a grid based finite volume scheme, with a High
Resolution Shock Capturing (HRSC) method.\footnote{Note that, even though
we may expect smooth steady state solutions based on the analytical results, we
are exploring an {\it a priori} unknown scenario in which shock fronts might
develop during the evolution, or even persist in the stationary state.}
The set of equations are written in a conservative form using
a variation of the ``3+1 Valencia formulation''~\cite{fBjaFjmIjmMjaM97} for
time independent, fixed metrics~\cite{lDZoZnBpL07}. The time integration is
achieved by adopting a second order total variation diminishing Runge-Kutta
method~\cite{cwSsO88}. The fluid evolution is performed in a fixed background
metric corresponding to a Kerr black hole using the same (horizon-penetrating)
Kerr-type coordinates adopted in Sec.~\ref{Sec:SteadyStateKerr}. The code
uses as primitive variables the rest-mass density, pressure and the locally
measured three-velocity vector $(\rho,P,v_i)$, where $v_i= \gamma_{ij} v^j$ and
\begin{equation}
   v^i = \frac{U^i}{\alpha U^t} + \frac{\beta^i}{\alpha},\qquad
  i = r,\theta,\phi,
   \label{eq:vi}
\end{equation}
with $\alpha$, $\beta^i$ and $\gamma_{ij}$ the lapse, shift vector and
three-metric of the 3+1 formalism~\cite{mA08}, written in these coordinates.
See~\cite{aguayo18,eTaA19,aAeTxH19,TAH20}, for more details about the
characteristics, test suite, and discretization method of \textsc{aztekas}.

For all the simulations presented in this section, we adopt an axisymmetric 2D
numerical domain $(r,\theta) \in [\mathcal{R}_{\mathrm{acc}},\mathcal{R}]\times
[0,\pi/2]$, with a uniform polar grid and an exponential radial grid (see
\cite{aAeTxH19} for details), where $\mathcal{R}$ is the radius of the outer
boundary at which we implement a free outflow condition for the velocities
and a fixed profile for the density and pressure. The inner boundary, set at
$\mathcal{R}_{\text{acc}} = 1.1 \,M$, for which we impose free outflow in all
the variables, is chosen such that $r_- < \mathcal{R}_{\mathrm{acc}} < r_+$
for all the explored values of $a$.  We fix reflection conditions at both
polar boundaries. A dissipative, second-order piecewise linear reconstruction
for the primitive variables is used in order to avoid spurious oscillations
due to these fixed boundary conditions.

In all the simulations, we evolve the equations from an initial state
consisting of a constant density and pressure gas cloud, with zero initial
three-velocity $v_i = 0$. The convergence to a steady-state is monitored
by computing the mean mass accretion rate $\dot{M}$ all over the domain,
until its variation drops below 1 part in $10^{4}$.

\subsection{Benchmark test}
\label{subsec:benchmark}

Taking advantage of the exact analytic description presented in the previous
sections, we use the solution in \eq{s3e1} as a benchmark test to prove
the convergence and stability of the \textsc{aztekas} code for this type of
problems. Moreover, this test is important in order to validate the subsequent
simulations discussed in this article. For these tests we implement the
ultrarelativistic stiff equation of state in the numerical code.

We reproduce the analytic solution corresponding to the choked accretion
model with $\mathcal{R} = 10M$, $V_0 = 0.2$ and a black hole spin $a=0.99M$.
We run the simulations in units such that $M=1$ and set the value  $\rho_0 =
1$ for the density at the reference point, although we remark here that, just
as in the analytic case, the resulting steady-state solution is scale-free
with respect to this specific value of $\rho_0$.  We perform four tests
varying the spatial resolution by a factor of 2 each time (with number of
grid points in the radial and polar directions $64\times64$, $128\times128$,
$256\times256$, $512\times512$, respectively). The values for $(\rho,P,v_i)$
from the analytic solution are imposed at the injection sphere as the boundary
condition, and these values are extended into the whole numerical domain as
the initial condition.

In Fig.~\ref{fig:comp-analytic} we show the isocontour levels of the density
field and of the magnitude of the three-velocity $V$ (as measured by a ZAMO)
of the \textsc{aztekas} simulations when the steady-state is reached at
$t=180\,M$. Likewise, the streamlines of the stationary flow are shown
in both figures. From these simulations we obtain the stagnation point
at $\mathcal{S} \simeq 4.6015 \,M$ which coincides with the analytical value
within the resolution uncertainty (see Fig.~\ref{fstiff}).

\begin{figure}[t]
 \centering
 \includegraphics[width=0.48\textwidth]{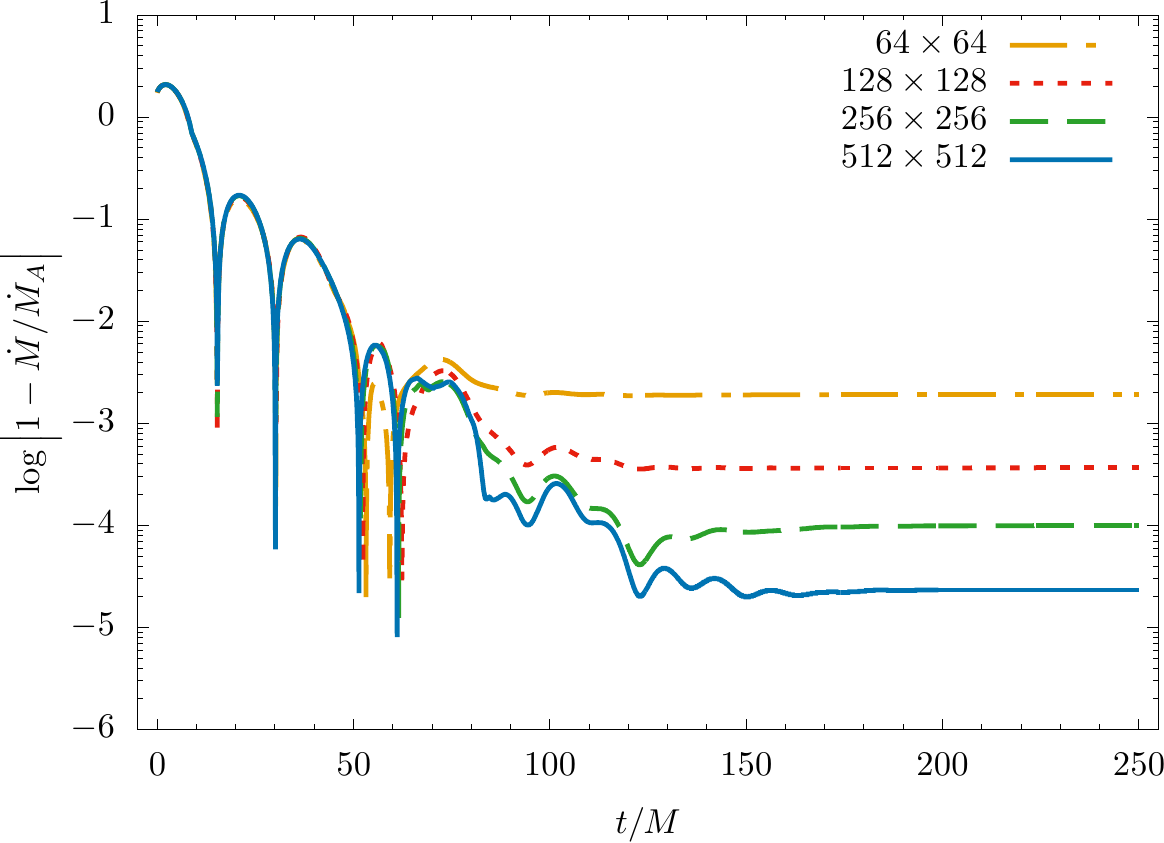}
 \caption{Validation test for the \textsc{aztekas} code. In this figure we
 show the evolution in time of the relative error between the numerical mass
 accretion rate $\dot{M}$ and analytic value $\dot{M}_A$ for the solution
 with parameters $\mathcal{R} = 10\, M$, $V_0 = 0.2$, and $a=0.99M$. The
 four different resolutions used for this benchmark test are represented
 with different dashed lines, showing a diminishing of the error as the
 resolution increases which, as shown in Appendix~\ref{App:convergence},
 is consistent with second order convergence.}
 \label{fig:mdot}
\end{figure}

In Fig.~\ref{fig:mdot} we show the evolution in time of the relative
error between the numerical mass accretion rate $\dot{M}$ and the analytic
value $\dot{M}_\mathrm{A}$, for all of the resolutions considered here. As
expected, the relative error decreases for larger resolutions.
Indeed, as further shown in Appendix~\ref{App:convergence},
from this benchmark test we confirm a second order convergence rate, as
expected from the adopted numerical scheme.

\subsection{Polytropic fluid}
\label{subsec:poly}

\begin{figure*}
 \centering
 \includegraphics[width=0.49\textwidth]{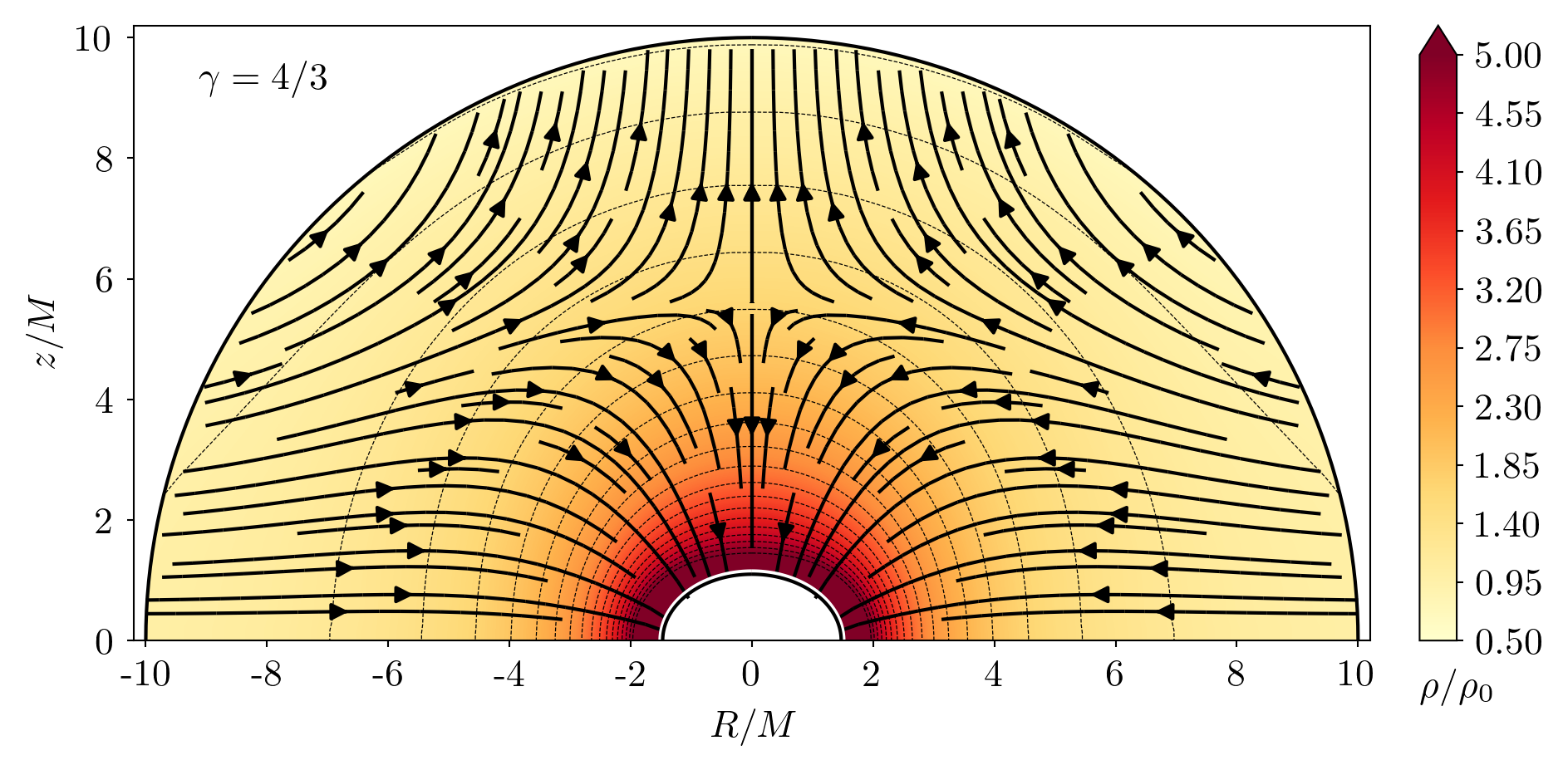}
 \includegraphics[width=0.49\textwidth]{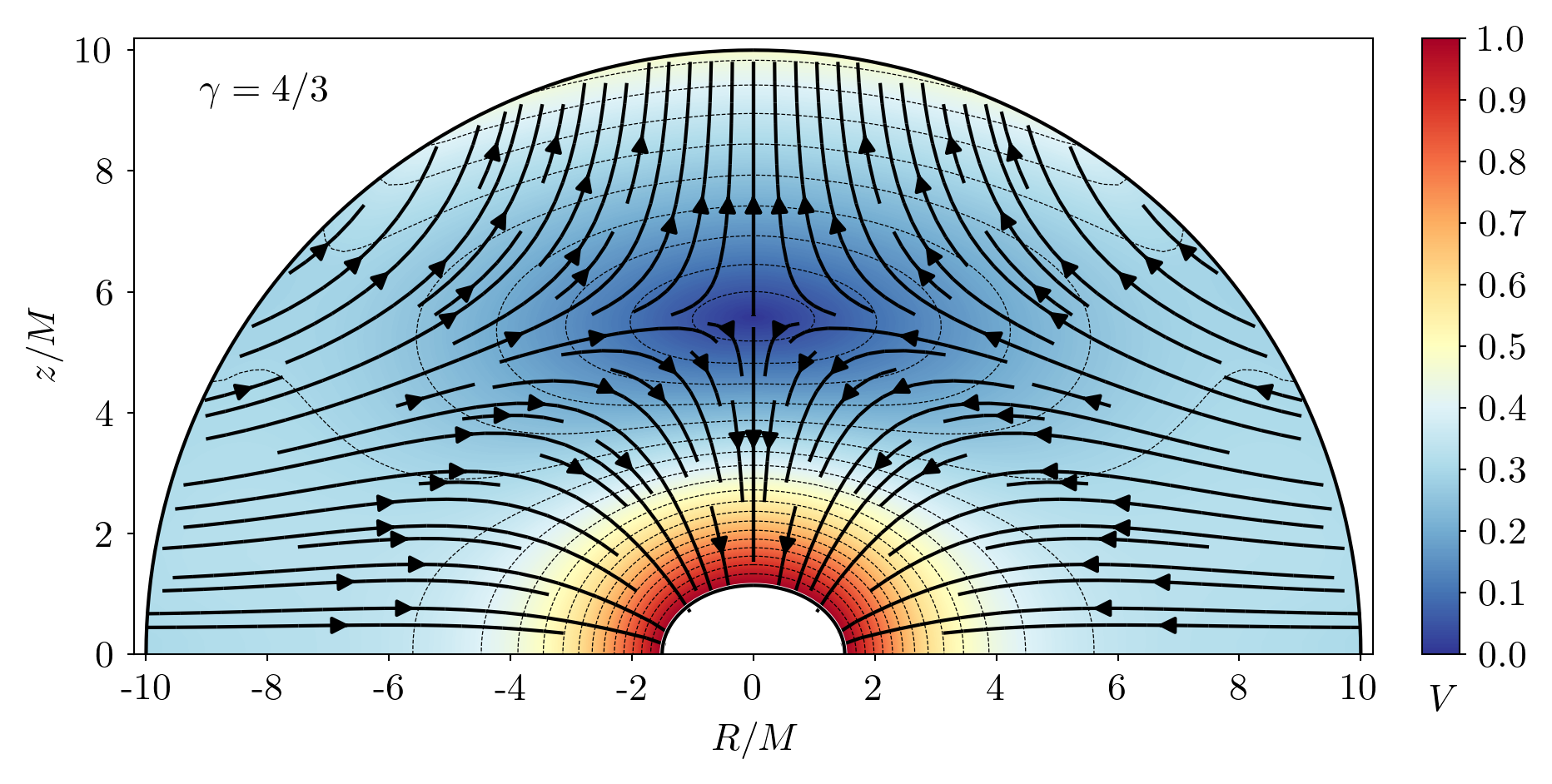}
 \includegraphics[width=0.49\textwidth]{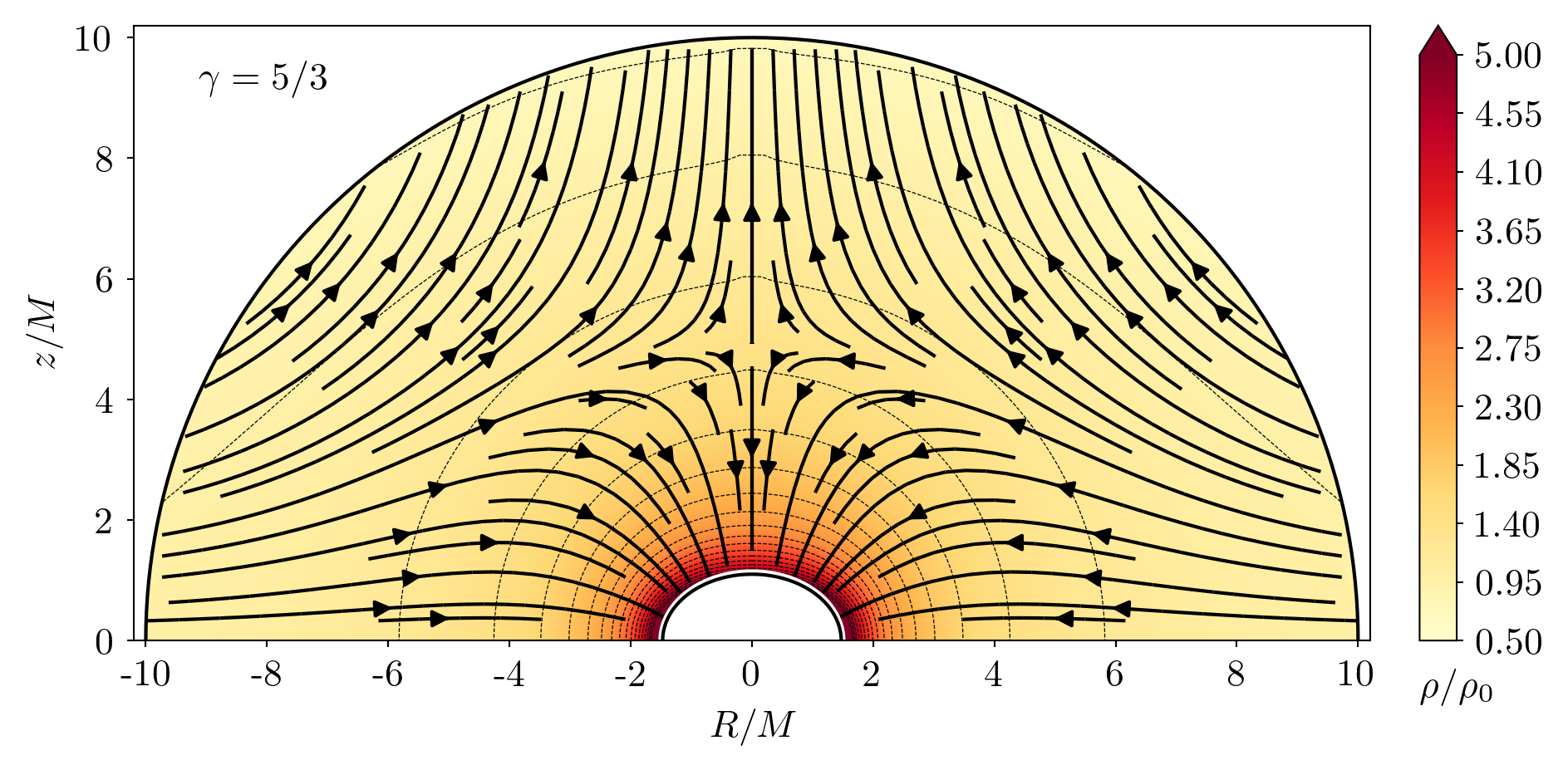}
 \includegraphics[width=0.49\textwidth]{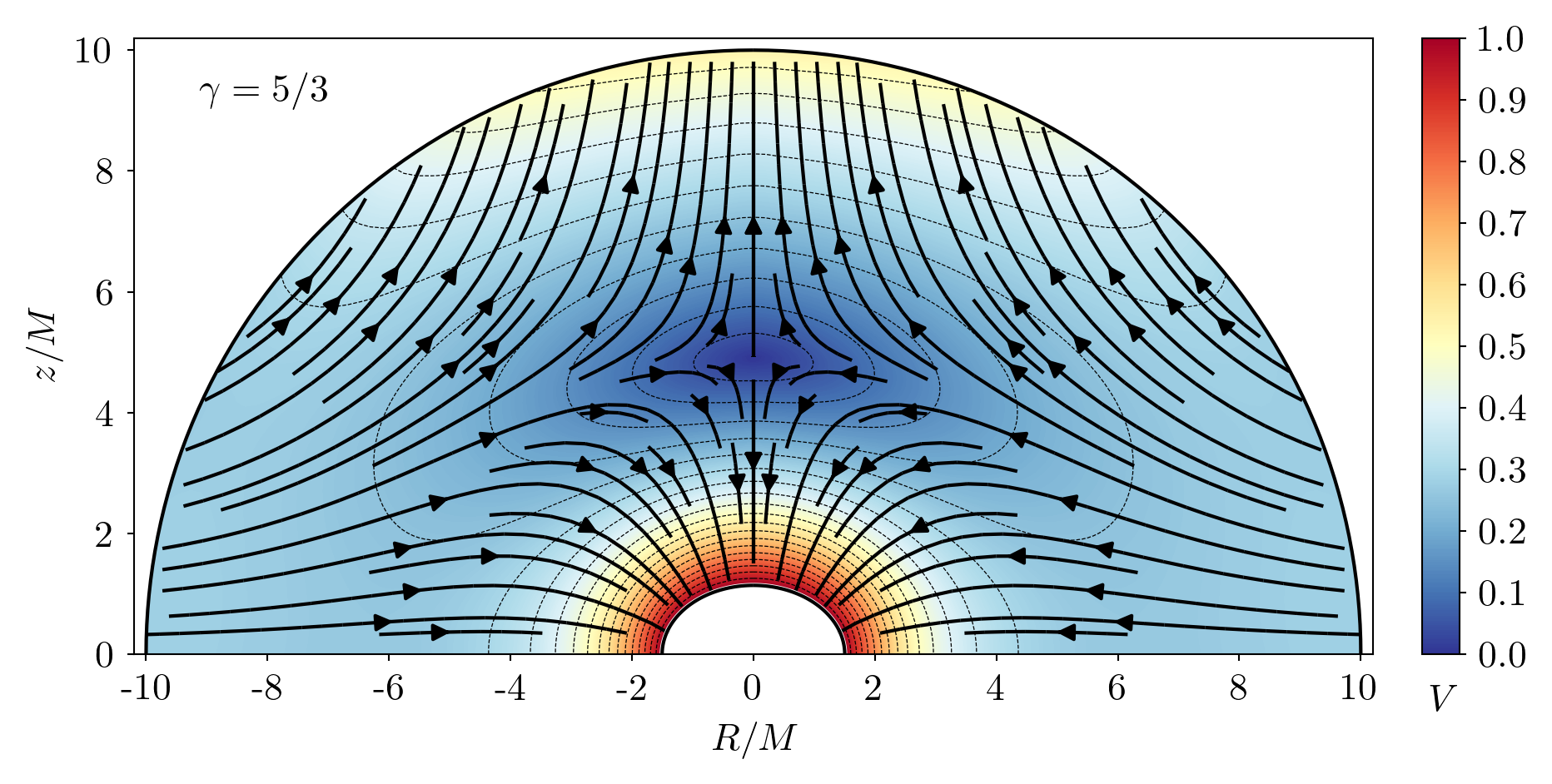}
 \caption{Stationary state of the numerical simulations of the choked accretion
 mechanism for a rotating black hole with $a=0.99M$, an injection radius of
 $\mathcal{R} = 10M$, a dimensionless temperature of $\Theta_0 =1$ at the
 equator of the injection sphere, and a polytropic fluid with $\gamma = 4/3$
 (top panels) and $5/3$ (bottom panels). On the left panels of this figure we
 show the isocontour levels of the normalized rest-mass density $\rho/\rho_0$
 while, on the right panels, the magnitude of the three-velocity $V$ as
 measured by the ZAMO.  The fluid streamlines are indicated with black
 solid arrows, while a white solid line shows the location of the event
 horizon $r_+$.}
 \label{fig:num_plots}
\end{figure*}

In order to explore the behavior of the choked accretion mechanism for a gas
with a less restrictive equation of state, we perform numerical simulations of
an ideal gas with a polytropic relation $P = K\rho^\gamma$, where $\gamma$
is the adiabatic index. We run experiments using a $256\times 256$ grid
resolution, for a wide range of values of the spin parameter $a/M \in [0,1)$
and two different values of the adiabatic index $\gamma \in \left\lbrace
4/3,5/3 \right\rbrace$.

The main feature of the choked accretion mechanism relies on the existence of
a density contrast at the external boundary. Following closely the boundary
treatment of~\cite{aAeTxH19,TAH20}, we fix the gas rest-mass density at the
outer boundary as
\begin{equation}
   \rho_{\mathrm{b}}(\theta) = \rho_0 (1 - \delta \cos^2 \theta),
   \label{eq:rho_b}
\end{equation}

\noindent where $\rho_0$ is the rest-mass density at the reference point
$(\mathcal{R},\pi/2)$ and $\delta$ is the density contrast defined as
\begin{equation}
   \delta = 1 - \frac{\rho_b(0)}{\rho_0}.
   \label{eq:delta}
\end{equation}

As mentioned in Sec.~\ref{Sec:introduction}, this density profile is motivated
as a way to introduce the axisymmetric anisotropy associated with fluid
rotation. In particular, it has been shown that low angular momentum fluids
accreting onto a central massive object give rise to a quasi-spherical, oblate
density distribution, as long as the angular momentum is sufficiently low
as to avoid encountering the centrifugal barrier \citep{proga2003,mach2018}.

The pressure at this boundary is then determined by the polytropic relation $P = K
\rho^\gamma$, where $K$ is computed as \cite{eTaA19}
\begin{equation}
   K = \frac{1}{\rho_0^{\gamma - 1}} \left[ \frac{c_0^2(\gamma -
   1)}{\gamma(\gamma - 1) - c_0^2} \right]
   \label{}
\end{equation}
with $c_0$ the speed of sound at the reference point.

An extensive exploration of the choked accretion mechanism's dependence on
$\mathcal{R}$, $\delta$ and $c_0$ can be found in~\cite{aAeTxH19} for the
non-relativistic regime and in~\cite{TAH20} for the case of a Schwarzschild
black hole. We performed a quick exploration of these three parameters,
for a rotating black hole with $a = 0.99M$, and found essentially the same
results as reported in those previous works. Moreover, we noticed that it
is more intuitive, in order to compare with possible astrophysical settings,
to use the dimensionless temperature\footnote{Note that our definition
of the dimensionless temperature is only valid for an ideal gas equation
of state. In terms of natural units, the general definition of the dimensionless temperature~\citep{mAcF2013} is
\mbox{$\Theta = k_\mathrm{B}T/m_\mathrm{b}c^2$},
with $T$ the fluid temperature, $c$ the speed of light, $k_\mathrm{B}$
Boltzmann's constant, and $m_\mathrm{b}$ the average baryonic mass.}
$\Theta_0 = P_0/\rho_0$ rather
than specifying $c_0$. For this reason, in what follows, we shall take as
representative values $\mathcal{R} = 10M$ for the domain size, $\delta =
0.5$ for the density contrast and $\Theta_0 = 1$ for the temperature of the
gas at the reference point. This value of the dimensionless temperature
corresponds to $c_0 \approx 0.52$ and $c_0 \approx 0.69$, for $\gamma =
4/3$ and $5/3$, respectively.  Furthermore, in order to have an appropriate
baseline reference for each combination of the $\gamma$ and $a$ parameters,
we also run simulations corresponding to ``spherical'' accretion in each case
(i.e., same values for $\gamma$ and $a$ but a $\delta = 0$ density contrast).

We evolve all the simulations until the stationary state has been reached
(within the previously mentioned limit of accuracy in which variations in the
mean accretion rate drop below $1$ part in $10^4$). The relaxation time depends
on $\gamma$, as well as on the value of $a$, but in all cases it is found
to conform to $500\, M < t < 1500\, M$. We also perform a self-convergence
test which is presented in Appendix~\ref{App:convergence}.

In Fig.~\ref{fig:num_plots} we show the resulting steady-state, rest-mass
density field and magnitude of the three-velocity $V$ for the $a=0.99\,
M$ case. The top panels show the results corresponding to $\gamma = 4/3$
while the bottom panels those for $\gamma = 5/3$. The black solid arrows
represent the fluid streamlines and the solid white line the location of
the outer horizon $r_+$.  As we can see from these figures, there is not a
strong qualitative difference in the flow morphology for different values
of $\gamma$, neither for the one presented in the non-rotating black hole
case~\cite{TAH20}. Moreover, although the streamlines configuration are
similar to the analytical case, the ejection velocity at the polar region
is larger for the polytropic fluid (see Fig.~\ref{fstiff}).

\begin{figure}
 \centering
 \includegraphics[width=0.45\textwidth]{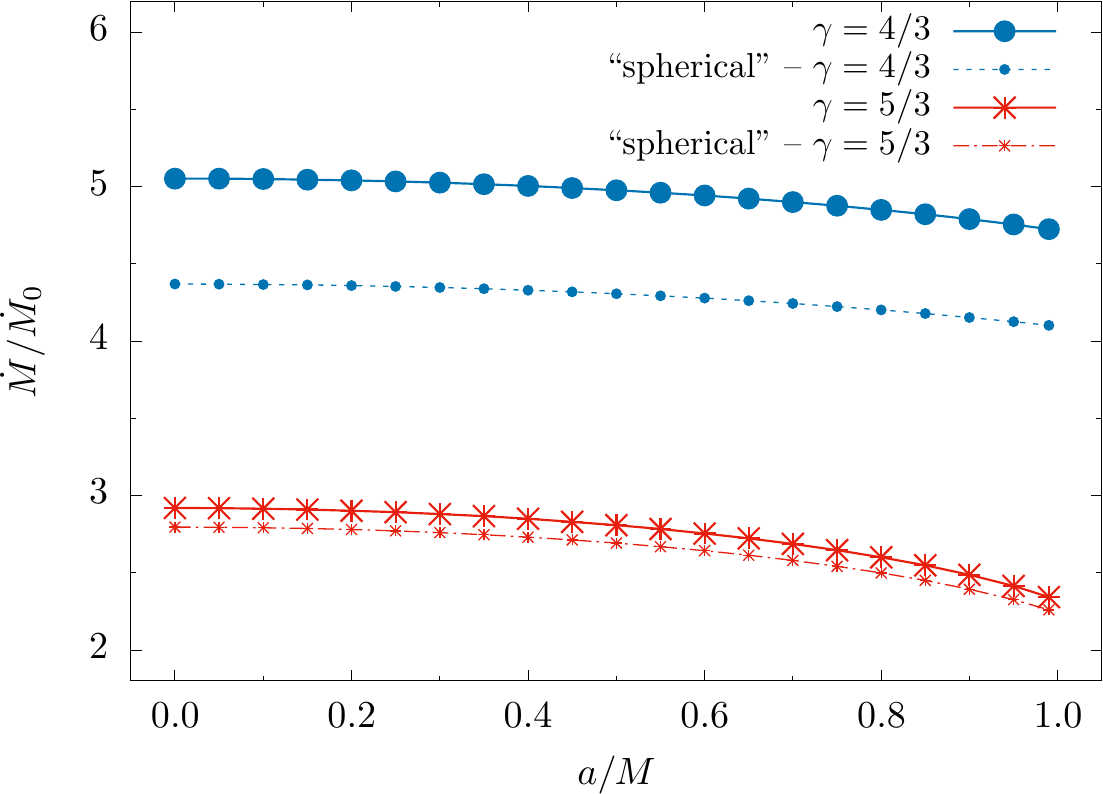}
 \caption{Mass accretion rate as a function of the spin parameter. The blue
 and red dotted lines correspond to $\gamma = 4/3$ and $5/3$, respectively. The
 black dashed lines represent their respective ``spherical'' accretion values,
 i.e., $\delta = 0$.}
 \label{fig:mdot-comp}
\end{figure}

In Fig.~\ref{fig:mdot-comp} we show the dependence of the mean mass accretion
rate $\dot{M}$ on the spin parameter $a$, for all the simulations performed
in this study. The blue dots correspond to $\gamma = 4/3$, while the red
crosses to $\gamma = 5/3$. The points joined by the dashed lines represent
the corresponding ``spherical'' accretion case ($\delta = 0$). It is
interesting to notice from this figure that, for each value of $\gamma$,
the dependence on $a$ remains the same regardless of the value of $\delta$
(except for a re-scaling factor that depends on $\gamma$).  This suggest
that the change in the mass accretion rate with the spin parameter is an
intrinsic characteristic of the accretion onto a Kerr black hole, and not of
the choked accretion mechanism.  A more complete study of the ``spherical''
accretion case onto a rotating black hole will be explored elsewhere.

In addition to the mass accretion rate, we compute the mass injection rate
$\dot{M}_\mathrm{in}$ and the mass ejection rate $\dot{M}_\mathrm{ej}$ at the
injection sphere (as defined in Sec.~\ref{sec:MinMej}). We also extract 
from the simulation's results the
location of the stagnation point $\mathcal{S}$. In Table~\ref{tab:tab1} we
present a summary of these results for a representative set of the performed
simulations. We also measure the magnitude of the three-velocity at the equator
$V_0$ and at the pole $V_\mathrm{ej}$, which do not present a significative
dependence on the spin parameter, maintaining a value around $V_0 = 0.30$ and
$V_\mathrm{ej} = 0.54$, for $\gamma = 4/3$; and $V_0 = 0.29$ and
$V_\mathrm{ej} = 0.48$, for $\gamma = 5/3$.

\begin{figure}[t]
 \centering
 \includegraphics[width=0.48\textwidth]{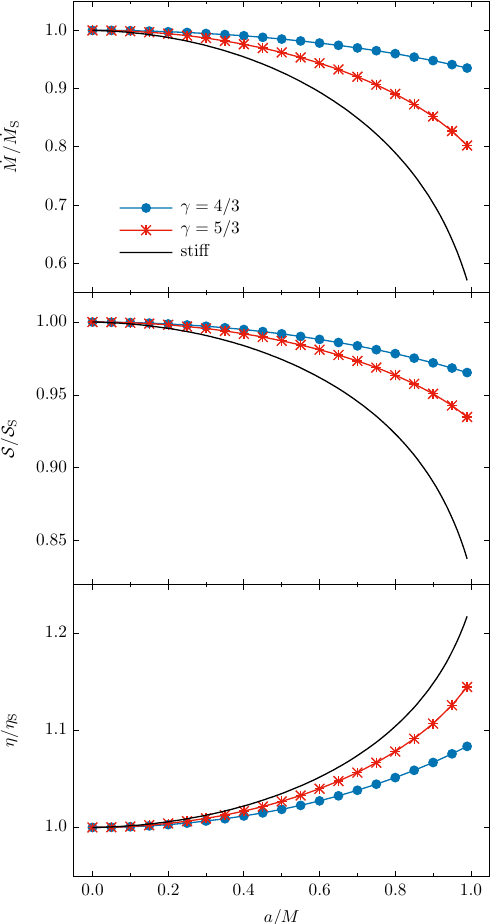}
 \caption{Dependence of different properties of the choked accretion
 simulations on the spin parameter $a/M$ and $\gamma$. From top to bottom, each
 panel shows: the mass accretion rate $\dot{M}$, the location of the stagnation
 point $\mathcal{S}/M$, and the ejection-to-injection mass rate ratio $\eta =
 \dot{M}_\mathrm{ej}/\dot{M}_\mathrm{in}$. Each panel is normalized by its
 corresponding non-rotating case value, which is denoted with the subscript
 $\mathrm{S}$. The black solid line represents the analytic solution presented
 in Sec.~\ref{Sec:choked} for a ultrarelativistic stiff fluid.}
 \label{fig:multi}
\end{figure}

In Fig.~\ref{fig:multi} we show the dependence on $a$ for the mass accretion
rate $\dot{M}$ (top panel), the location of the stagnation
point $\mathcal{S}$ (middle panel), and the
ejection-to-injection mass rate $\eta$ (bottom panel), for both values of
$\gamma$. In order to clearly see
the change with $a$ for these quantities, we normalize them by
$\dot{M}_\mathrm{S}$, $\mathcal{S}_\mathrm{S}$ and $\eta_\mathrm{S}$,
respectively, which correspond to the values obtained in the non-rotating case
($a=0$). Moreover, we also include the stiff analytic solution obtained in
Sec.~\ref{Sec:choked}, using as the $V_0$ parameter the value found for $\gamma
= 5/3$.

\begin{table*}
\setlength{\tabcolsep}{17pt}
\caption{Results for our simulations with parameters $\mathcal{R} = 10\, M$, $\delta =
0.5$ and $\Theta_0 = 1$. }
\begin{tabular}{c|cccc|cccc}
\hline
   \multicolumn{1}{c}{}
  & \multicolumn{4}{c}{$\gamma = 4/3$} &
  \multicolumn{4}{c}{$\gamma = 5/3$} \\
\hline
\hline
$a/M$ & 
$\mathcal{S}/M$ & $\dot{M}$ &
$\dot{M}_\mathrm{ej}$ & $\dot{M}_\mathrm{in}$ &
$\mathcal{S}/M$ & $\dot{M}$ &
$\dot{M}_\mathrm{ej}$ & $\dot{M}_\mathrm{in}$ \\
\hline
0.0  & 5.781 & 5.052 & 2.195 & 7.246 & 5.247 & 2.919 & 2.768 & 5.687 \\
0.25 & 5.770 & 5.033 & 2.201 & 7.234 & 5.231 & 2.892 & 2.777 & 5.670 \\
0.5  & 5.734 & 4.977 & 2.220 & 7.197 & 5.179 & 2.808 & 2.805 & 5.614 \\
0.75 & 5.672 & 4.876 & 2.256 & 7.132 & 5.083 & 2.647 & 2.857 & 5.504 \\
0.99 & 5.581 & 4.725 & 2.308 & 7.032 & 4.905 & 2.343 & 2.947 & 5.289 \\
\hline 
\end{tabular}\\
{NOTE -- The quantities $\dot{M}$, $\dot{M}_\mathrm{in}$ and
$\dot{M}_\mathrm{ej}$ are given in units of $\dot{M}_0 = 8\pi M^2 \rho_0$.}
\label{tab:tab1}
\end{table*}

As we can see from Fig.~\ref{fig:multi}, the mass accretion rate decreases with
the spin parameter down to a factor of $\sim10\%$ (20\%) for the $\gamma=4/3$
($\gamma=5/3$) case as the spin parameter increases to its maximum value. On
the other hand, the location of the stagnation point only decreases down
to a factor of $\sim5$\% for both values of $\gamma$. In contrast, the
ejection-to-injection mass rate $(\eta)$ increases up to a factor of 10 to
15\% as $a\to M$.
 
Even though it is not possible to make a direct comparison of the numerical
simulations with the analytic model of an ultrarelativistic stiff fluid,
since in each case we are using different equations of state and the boundary
conditions are not exactly the same, there are still some observations that
can be drawn from Fig.~\ref{fig:multi}. First of all, there is a shared,
qualitatively consistent dependence of the different quantities shown in
this figure on the spin parameter $a$, both for the polytropic gas and the
stiff fluid. Moreover, it is also clear that there is a stronger response
from the stiff fluid to the black hole rotation. Indeed, from this figure
we see that the analytic solution presented in Sec.~\ref{Sec:choked} can be
used as a lower limit for the mass accretion rate and for the location of the
stagnation point that would follow for a polytropic gas, whereas it can be
used as an upper limit for the ejection-to-injection mass rate ratio. Finally,
we note  that there is a clear trend for a stronger dependence on the spin
parameter $a$ as the fluid stiffens ($\gamma\to2$).

\section{Summary and conclusions}
\label{Sec:Conclusions}

The choked accretion model is a purely hydrodynamical mechanism with which it
is possible to obtain a bipolar outflow by perturbing an originally radial
inflow. The necessary conditions for this mechanism to operate consist
of a sufficiently large mass accretion rate onto a central massive object
(as compared to the Bondi accretion rate), and an anisotropic density field
in which the equatorial region is at a higher density than at the poles.
Potential astrophysical applications of this model for outflow-generating
phenomena are mentioned in the introduction and have been discussed in
further detail in~\cite{aAeTxH19,TAH20}.

In this article we have presented a generalization of the choked accretion
mechanism for the case of a rotating Kerr black hole, extending the
perturbative study initiated by Hernandez et al.~\cite{hernandez14} and
the subsequent analytical and numerical studies at the non-relativistic
level~\cite{aAeTxH19} and in the Schwarzschild case~\cite{TAH20}.
Here we have shown, using both analytic solutions and numerical simulations,
that the choked accretion's main features are recovered in the presence of
a rotating black hole, regardless of the value of the spin parameter.

Our analytic model is based on the steady-state, irrotational solution
for an ultrarelativistic stiff fluid presented by Petrich, Shapiro
and Teukolsky~\cite{lPsSsT88}. We have derived the general equations
of the model using horizon-penetrating Kerr-type coordinates and
then mostly focused on the axisymmetric quadrupolar case, studying
the dependence of the flow morphology on the unique parameter $A$
that remains free (we have also briefly discussed the misaligned
quadrupolar case at the end of Sec.~\ref{Sec:AxisymmetricQuadFlow} and in
Appendix~\ref{App:Misaligned}). Depending on the sign of this parameter,
the flow describes an equatorial inflow-bipolar outflow solution ($A>0$)
or an equatorial outflow-polar inflow solution ($A<0$).  Given that it
has a wider applicability in an astrophysical context and corresponds to
the choked accretion scenario discussed in this article, we have mainly
focused on the case $A > 0$ and discussed the physical properties of the
choked accretion model, including its mass accretion rate, location of the
stagnation points and ejection-to-injection mass rate ratio.  We have also
extended the present study to a perfect fluid obeying a polytropic relation
with adiabatic index $\gamma=4/3$ and $5/3$, by performing full hydrodynamic,
relativistic numerical simulations of an ideal gas in a Kerr background metric.

In previous works, it was found that the total mass accretion rate in the
choked accretion model has a threshold value close to the one found in
the spherical accretion scenario. In this study, based on both analytic
and numerical analysis, we have extended this result to the case of a
rotating black hole and shown that the accretion rate obtained from the
``spherically symmetric'' case, in which the density contrast is set to
zero at the injection sphere, still yields a lower limit for the threshold
value for the choking mechanism to work (see Figs.~\ref{fig:analytic} and
\ref{fig:mdot-comp}). Note, however, that in the case of the Kerr spacetime
there is no analytic equivalent to Michel's solution in the Schwarzschild
case. Therefore, this problem has to be studied by numerical means as we
have briefly discussed here and will further address in a future work.

Most of the configurations analyzed in this article have focused on the
aligned case, in which the axis of the bipolar outflow coincides with the
rotation axis of the black hole, in which case the fluid elements have zero
angular momentum and hence would not affect the black hole's spin during
their accretion. However, in Appendix~\ref{App:Misaligned} we have also
analyzed a misaligned configuration, and it is interesting to note that such
an accretion flow would slow down the black hole's rotation, as the results in
Sec.~\ref{SubSec:Conservation} show.

This work continues a series of analytic and numerical studies of the
choked accretion model as a purely hydrodynamical mechanism for generating
axisymmetric outflows. In future work we intend to expand the ingredients
involved in this model, by including additional physics such as fluid angular
momentum, viscous transport, and magnetic fields, in order to explore the
applicability of the model in outflow-generating astrophysical systems.

\acknowledgments
It is a pleasure to thank Diego L\'opez-C\'amara and Xavier Hern\'andez for
fruitful discussions and useful comments on a previous version of this article.
We also thank the anonymous referee for helpful suggestions and remarks. This
work was supported in part by CONACyT (CVUs 788898, 673583) and by a CIC Grant
to Universidad Michoacana.

\appendix

\section{The misaligned quadrupolar flow}
\label{App:Misaligned}

In Sec.~\ref{Sec:AxisymmetricQuadFlow} we discussed the axisymmetric
quadrupolar flow solution and some important properties regarding its
morphology, and in Sec.~\ref{Sec:choked} this solution was applied
to the choked accretion scenario. This flow has the property of being
reflection-symmetric about the equatorial plane of the Kerr black hole,
such that the bipolar outflow regions are aligned with the symmetry
axis. In this appendix, we discuss an example in which the flow discussed in
Secs.~\ref{Sec:AxisymmetricQuadFlow} and \ref{Sec:choked} is ``rotated"
by an angle $\theta_0$ about an axis within the plane $\theta = \pi/2$
(in a sense made precise below).

When the black hole is non-rotating, the aforementioned rotation can be carried
out exactly and is simply a rigid rotation of the (spherically symmetric)
Schwarzschild geometry the Kerr metric reduces to in the limit $a = 0$. We may
construct this rotated solution explicitly by writing the angular dependency
in the axisymmetric quadrupolar flow solution~(\ref{s3e2}) in the form
\begin{equation}
3\cos^2\theta - 1 = 2P_2(\cos\theta) = 2P_2({\bf x}'\cdot {\bf x}),
\label{Eq:P2}
\end{equation}
with $P_2$ denoting the Legendre polynomial $P_\ell$ with $\ell=2$ and
${\bf x} =  (\sin\theta\cos\phi,\sin\theta\sin\phi,\cos\theta)$ and ${\bf
x'} := (0,0,1)$. Applying a rotation by the angle $\theta_0$ about the $y$
axis is equivalent to replacing the vector ${\bf x}' = (0,0,1)$ with the
vector ${\bf x}' = (\sin\theta_0,0,\cos\theta_0)$ in the right-hand side
of \eq{Eq:P2}. Recalling the addition theorem for spherical harmonics (see,
for instance, chapter~3.6 in Ref.~\cite{Jackson-Book})
\begin{equation}
P_\ell({\bf x}'\cdot {\bf x}) = \frac{4\pi}{2\ell + 1}\sum\limits_{m=-\ell}^\ell 
(Y^{\ell m})^*(\theta',\phi') Y^{\ell m}(\theta,\phi),
\label{Eq:SphHarmonicAdditionThm}
\end{equation}
we can write the rotated quadrupolar flow solution on a Schwarzschild
background as in \eq{s3e1} with $a = 0$ and $F(r,\theta,\phi)$ given by
\begin{equation}
F(r,\theta,\phi) = \frac{8\pi}{5}(3r^2 - 6Mr + 2M^2)\sum\limits_{m=-\ell}^\ell d_m Y^{\ell m}(\theta,\phi)
\label{Eq:FRotated}
\end{equation}
with the coefficients $d_m := (Y^{\ell m})^*(\theta_0,0)$. This has again the
form of the general solution in \eq{Eq:PSTSolution} when $a = 0$, and hence it
describes a solution of the potential flow equation~(\ref{Eq:Wave}). However,
it bears exactly the same physical content as the original axisymmetric
quadrupolar flow solution discussed in Secs.~\ref{Sec:AxisymmetricQuadFlow}
and \ref{Sec:choked}, since it is obtained from it by an isometry.

When $a\neq 0$, the method we have just described cannot be
performed, since merely replacing ${\bf x}' = (0,0,1) \mapsto {\bf x}' =
(\sin\theta_0,0,\cos\theta_0)$ in the right-hand side of \eq{Eq:P2} would not
yield a solution of \eq{Eq:Wave}. This is due to the $m$-dependency in the
radial functions appearing in the expansion~(\ref{Eq:PSTExpansion}) which,
in turn, arises because of the lack of spherical symmetry of the Kerr metric
when $a\neq 0$. On the other hand, we still have the freedom of choosing the
five complex constants $A_{2 m}$ in \eq{Eq:PSTExpansion}, as long as they
satisfy the reality conditions~(\ref{Eq:RealityConditions}). In particular,
we can choose these coefficients such that the function $F(r,\theta,\phi)$
has the same weights $d_m$ as in \eq{Eq:FRotated} on \emph{some particular
constant $r$ surface}. This is equivalent to applying the rotation ${\bf x}' =
(0,0,1) \mapsto {\bf x}' = (\sin\theta_0,0,\cos\theta_0)$ on this particular
surface only, which yields
\begin{eqnarray}
&& F(r,\theta,\phi) = \frac{4\pi}{5}(r_+ - r_-)^2 F(-2,3;1,-x_*)
\nonumber\\
&& \times \sum\limits_{m = -2}^2 \frac{F(-2,3; 1+i\,m\,\alpha, -x)}{F(-2,3; 1+i\,m\,\alpha, -x_*)} 
d_m Y^{2 m}(\theta,\phi),\qquad
\label{Eq:F2m}
\end{eqnarray}
where we recall that $\alpha = 2a/(r_+ - r_-)$ and $x = (r - r_+)/(r_+ -
r_-)$ and $x_*$ is the value of $x$ corresponding to the location of the
surface where the rotation is applied. Note that for $x = x_*$ the weight
functions are the same as in \eq{Eq:FRotated}, as required. Notice also
that when $\theta_0 = 0$ (in which case only the $m = 0$ mode contributes
and $(Y^{20})^*(0,0) Y^{20}(\theta,\phi) = 5P_2(\cos\theta)/(4\pi)$),
the function $F$ in \eq{Eq:F2m} reduces to the function $F$ in \eq{s3e2}
in the aligned case.

To determine uniquely the solution it remains to choose the value for
$x_*$. One possibility is to choose it such that it corresponds to the radius
of the injection sphere $r = {\cal R}$. However, note that for finite $r$,
the two-surfaces $(t,r) = \co$ are not strict metric spheres in the Kerr
geometry, so we argue that only in the asymptotic limit $r\to \infty$ does
it make sense to apply the rotation in a sensible way. Hence, even though
the flow is only well-defined inside a finite region, we exploit the fact
that the potential $\Phi$ itself \emph{is} well-defined for all $r > r_-$
and thus we take the limit $x_*\to \infty$ in \eq{Eq:F2m}. This finally yields
\begin{eqnarray}
&& F(r,\theta,\phi) = \frac{4\pi}{5}\sum\limits_{m = -2}^2\left[ 6r^2 - 6(2M - i\,m\,a) r  \right. 
\nonumber\\
&& \left. + 4M^2 + 2a^2 - 2m^2 a^2 - 6\,i\,m\,a\,M \right] d_m Y^{2 m}(\theta,\phi).\qquad
\label{Eq:F2mBis}
\end{eqnarray}
Using the explicit representation of the spherical harmonics one can write
the result in the form
\begin{eqnarray}
&& F(r,\theta,\phi) = \frac{1}{4} a_0(r)(3\cos^2\theta_0-1)(3\cos^2\theta - 1)
\nonumber\\
 &&+ 3 [a_1(r)\cos\phi + b_1(r)\sin\phi]\cos\theta_0\sin\theta_0\cos\theta\sin\theta
\nonumber\\
 &&+ \frac{3}{4} [a_2(r)\cos(2\phi) + b_2(r)\sin(2\phi)]\sin^2\theta_0\sin^2\theta,
\label{Eq:Misaligned1}
\end{eqnarray}
with the radial functions
\begin{eqnarray}
a_m(r) &=& 6r^2 - 12M r + 4M^2 + 2(1 - m^2)a^2,\\
b_m(r) &=& -6\,m\,a(r - M).
\end{eqnarray}

Another useful representation of the solution is obtained by writing it in
terms of the ``rotated" Cartesian coordinates $(r\xi,r\eta,r\zeta)$, where
\begin{equation}
\left( \begin{array}{c} \xi \\ \eta \\ \zeta \end{array} \right)
 = \left( \begin{array}{ccc} 
 \cos\theta_0 & 0 & -\sin\theta_0 \\ 
 0 & 1 & 0 \\
 \sin\theta_0 & 0 & \cos\theta_0 \end{array} \right)
 \left( \begin{array}{c} \sin\theta\cos\phi \\ \sin\theta\sin\phi \\ \cos\theta \end{array} \right).
\label{Eq:Rotation}
\end{equation}
This gives
\begin{eqnarray}
&& F(r,\theta,\phi) = 2\left( 3r^2 - 6M r + 2M^2 + a^2 \right) P_2(\zeta) 
\nonumber\\
&& - 6\,\varepsilon\left[ 3(r-M)\eta + a\cos\theta_0\xi \right]\zeta
 + 6\,\varepsilon^2(\eta^2 - \zeta^2),\qquad\quad
\label{Eq:Misaligned2}
\end{eqnarray}
with $\varepsilon := a\sin\theta_0$. Note that for $\varepsilon = 0$ (which is
the case if the black hole is non-rotating or the inclination angle $\theta_0$
vanishes), the second line in Eq.~(\ref{Eq:Misaligned2}) vanishes and one
recovers the axisymmetric quadrupolar flow solution~(\ref{s3e2}) with the
rotated symmetry axis $r\zeta$.

We conclude this appendix by showing that for small values of $|\varepsilon|$
the solution~\eq{s3e1} with $A > 0$ and $F$ as in~\eq{Eq:Misaligned2} still
has two stagnation points whose location can be determined by a perturbative
method. To this purpose we introduce the vector-valued function $H(\varepsilon;
w)$ with $w = (r,\xi,\eta)$, defined as
\begin{equation}
H(\varepsilon; w) := \left( \Delta\frac{\partial F}{\partial r} - \frac{2Mr_+}{A},
\frac{\partial F}{\partial\xi},\frac{\partial F}{\partial\eta} \right),
\end{equation}
where the constraint $\zeta^2 = 1 - \xi^2 - \eta^2$ should be taken
into account (since the function $F$ is symmetric with respect to
$(\xi,\eta,\zeta)\mapsto -(\xi,\eta,\zeta)$ it is sufficient to perform the
analysis for the case $\zeta > 0$). The location of the stagnation points
(for a given value of $\varepsilon$) is characterized by a zero of the
function $H(\varepsilon;\cdot)$, see~\eq{Eq:ZAMORest}. For $\varepsilon =
0$ one can check that the zero lies at
\begin{equation}
w = w_0 = (r_0,0,0),
\end{equation}
with $r_0 = {\cal S}$ as in~\eq{eA1}. To determine the location
$w(\varepsilon)$ of the zero for small values of $|\varepsilon|$ one can
differentiate both sides of the equation $H(\varepsilon; w(\varepsilon)) =
0$ with respect to $\varepsilon$, which gives
\begin{equation}
DH(\varepsilon; w(\varepsilon)) \frac{dw}{d\varepsilon}(\varepsilon) 
 + \frac{\partial H}{\partial \varepsilon}(\varepsilon; w(\varepsilon)) = 0,
\label{Eq:Diff}
\end{equation}
where $DH$ refers to the Jacobi matrix of $H$ with respect to $w$. Evaluating
at $\varepsilon = 0$ yields
\begin{equation}
DH(0; w_0) w_1 = -\frac{\partial H}{\partial \varepsilon}(0; w_0),\qquad
w_1 := \frac{dw}{d\varepsilon}(0).
\end{equation}
Since $DH(0; w_0) = 6K\diag(2,-1,-1)$ with $K := 3r_0^2 - 6M r_0 + 2M^2
+ a^2 > 0$, the first-order correction $w_1$ is uniquely determined
by this equation (and according to the implicit function theorem, the
function $H(\varepsilon;\cdot)$ has a unique zero for small enough values
of $|\varepsilon|$). By further differentiation of~\eq{Eq:Diff} one can
compute the higher-order corrections of $w(\varepsilon)$. Up to terms of
order $\varepsilon^3$ this gives
\begin{eqnarray}
r(\varepsilon) &=& r_0 + \frac{M r_+}{4A K^2}(3M^2 - 2a^2)\varepsilon^2 
 + {\cal O}(\varepsilon^3),\qquad
\label{Eq:reps}\\
\xi(\varepsilon) &=& -\frac{a}{K}\varepsilon + {\cal O}(\varepsilon^3),
\label{Eq:xieps}\\
\eta(\varepsilon) &=& -\frac{3(r_0-M)}{K}\varepsilon + {\cal O}(\varepsilon^3).
\label{Eq:etaeps}
\end{eqnarray}

A few numerical examples for the case $A M = 0.01$ and $a/M = 0.5$ are given in
Table~\ref{tab:stagnation}. The particular entry corresponding to $\theta_0 =
30^\circ$ corresponds to the flow shown in Fig.~\ref{fig:misaligned}.

Finally, we point out that the characterization of the stagnation point
we have used so far, based on the vanishing of the ZAMO's three-velocity,
might not be the most adequate definition from a conceptual point of view
when $\varepsilon \neq 0$. This is due to the fact that a ZAMO rotates
around the black hole with angular frequency $\Omega$ (see \eq{Eq:ZAMO}),
and hence such an observer which is located at $r = r(\varepsilon)$ and
$\theta = \theta(\varepsilon)$ only sees the fluid at rest in its frame at
the moments it crosses the plane $\phi = \phi(\varepsilon)$. In other words,
the world line of the stagnation point defined in this way does not agree
with the one of the ZAMO. An alternative definition of the stagnation point
which does not suffer from this problem can be given by requiring the fluid's
three-velocity of a static observer (as opposed to a ZAMO) to vanish. The
location of this point can be determined perturbatively by the same method
as the one we have just described; however we adopt the former definition
in view of the compatibility with Fig.~\ref{fig:misaligned} in which the
ZAMO's three-velocity is shown.

\begin{table*}
\caption{Location of the stagnation point for the parameter values $A M = 0.01$ and $a/M = 0.5$. Five significant figures are shown. The perturbative calculation refers to the expansion~(\ref{Eq:reps}--\ref{Eq:etaeps}), truncating the ${\cal O}(\varepsilon^3)$ terms and translated back to the angle coordinates $\theta(\varepsilon)$ and $\phi(\varepsilon)$ by means of Eq.~(\ref{Eq:Rotation}). The numerical calculation is based on the \mbox{fsolve} routine in MAPLE, using $15$ digits of precision and the seed values provided by the perturbative calculation. As can be appreciated from the table, the values provided by the quadratic expansion~(\ref{Eq:reps}--\ref{Eq:etaeps}) give a very good approximation (less than $1\%$ relative error in the quantities $(r(\varepsilon),\theta(\varepsilon),\phi(\varepsilon))$).}
\begin{tabular}{c|ccc|ccc}
\hline
   \multicolumn{1}{c}{} & \multicolumn{3}{c}{Perturbative calculation}
   & \multicolumn{3}{c}{Numerical calculation}  \\
\hline
\hline
$\theta_0$  & $r(\varepsilon)/M$ & $\theta(\varepsilon) - \theta_0$ & $\phi(\varepsilon)$
& $r(\varepsilon)/M$ & $\theta(\varepsilon) - \theta_0$ & $\phi(\varepsilon)$ \\
\hline
 $0^\circ$  & $4.2242$  & $0.0$ & $\hbox{(undefined)}$
                  & $4.2242$  & $0.0$ & $\hbox{(undefined)}$  \\
$10^\circ$ & $4.2243$ & $0.00073719$ & $-0.15890$      
                  & $4.2243$ & $0.00075882$ & $-0.15886$  \\
$20^\circ$  & $4.2244$ & $0.0012597$ & $-0.15902$        
                  & $4.2244$ & $0.0014289$ & $-0.15884$  \\
$30^\circ$  & $4.2245$ & $0.0013812$ & $-0.15919$     
                  & $4.2245$ & $0.0019307$ & $-0.15882$  \\               
$40^\circ$  & $4.2247$ & $0.00096893$ & $-0.15939$       
                  & $4.2247$ & $0.0022033$ & $-0.15879$ \\ 
$50^\circ$  & $4.2249$ & $-0.000035935$ & $-0.15959$           
                  & $4.2249$ & $0.0022116$ & $-0.15875$ \\
$60^\circ$  & $4.2250$ & $-0.0016067$ & $-0.15972$   
                  & $4.2251$ & $0.0019516$ & $-0.15872$  \\
$70^\circ$ & $4.2252$ & $-0.0036296$ & $-0.15979$
                 & $4.2253$ & $0.0014526$  & $-0.15870$ \\
$80^\circ$ & $4.2253$ & $-0.0059139$ & $-0.15974$ 
                  & $4.2254$ & $0.00077429$ & $-0.15868$
\end{tabular}
\label{tab:stagnation}
\end{table*}

\section{Bounds on the parameter $A$}
\label{App:Bounds}

In this appendix we prove that for sufficiently large radii ${\cal R}$ of
the injection sphere, the maximum range $A_-({\cal R}) < A < A_+({\cal R})$
for the parameter $A$ in the axisymmetric quadrupolar potential $\Phi$ in
\eq{s3e1} to yield a well-defined flow on the domain $r_+\leq r\leq {\cal R}$
is determined by the requirement for the magnitude of the three-velocity $V$
to be subluminal at the poles of the injection sphere. That is, we show that
for any large enough value of ${\cal R}$, $V < 1$ at the poles of the sphere
$r = {\cal R}$ guarantees that the gradient of $\Phi$ is everywhere timelike
on the domain $r_+\leq r\leq {\cal R}$.

To prove this claim, we go back to the investigation toward the end of
Sec.~\ref{Sec:AxisymmetricQuadFlow}, from which it follows that the gradient
$\nabla_\mu\Phi$ is timelike if and only if
\begin{equation}
\varrho^2\frac{h^2}{e^2} = c_2(r)\,\cos^4\theta + c_1(r)\,\cos^2\theta + c_0(r) > 0.
\label{Eq:h2Timelike}
\end{equation}
In the limit $A M\ll 1$ it was shown that the outer boundary of the
region for which~(\ref{Eq:h2Timelike}) holds describes a large ellipsoid
of revolution with semi-axes equal to $1/(6|A|)$, $1/(6|A|)$, $1/(12|A|)$
in the $x,y,z$-directions, respectively. It is then clear that for ${\cal
R}$ large enough, the outer boundary first intersects the sphere $r =
{\cal R}$ at the poles $\theta = 0,\pi$. The corresponding values for $A$
can be determined by evaluating the condition $c_2({\cal R}) + c_1({\cal R})
+ c_0({\cal R}) = 0$, which yields
\begin{equation}
A_\pm({\cal R}) = \pm\frac{1}{12({\cal R} - M)}
\left[ 1 + \frac{2M({\cal R}\pm r_+)} {({\cal R} - r_+)({\cal R} - r_-)} \right],
\label{Eq:Apm}
\end{equation}
and hence for large ${\cal R}$ the gradient $\nabla_\mu\Phi$ is timelike
on the sphere $r = {\cal R}$ if $A_-(R) < A < A_+(R)$. We now prove the
following statements, which show that these conditions are also sufficient
for the flow to be everywhere well-defined in the shell delimited by the
event horizon and the injection sphere.

\begin{theorem}
\begin{enumerate}
\item[(a)] Suppose ${\cal R} > r_+$ is large enough such that $12 r_+ A_+({\cal R}) \leq 1$, and let $0\leq A < A_+({\cal R})$. Then the right-hand side of \eq{Eq:h2Timelike} is strictly positive for all $r_+\leq r\leq {\cal R}$ and all $0\leq \theta\leq \pi$.
\item[(b)] Suppose ${\cal R} > r_+$ and $A_-({\cal R}) < A \leq 0$. Then the right-hand side of \eq{Eq:h2Timelike} is strictly positive for all $r_+\leq r\leq {\cal R}$ and all $0\leq \theta\leq \pi$.
\end{enumerate}
\end{theorem}

\proof For the proof it is convenient to rewrite the right-hand side of
\eq{Eq:h2Timelike} in the following form:
\begin{equation}
E_A(r,\xi) := d_2(r)\xi^2 + d_1(r)\xi + d_0(r),
\end{equation}
where $\xi := \sin^2\theta$ and the coefficients $d_0(r) := c_0(r) + c_1(r) +
c_2(r)$, $d_1(r) := -c_1(r) - 2c_2(r)$ and $d_2(r) := c_2(r)$ are explicitly
given by
\begin{subequations}
\begin{eqnarray}
d_0(r) &=& -144A^2\Delta(\Delta + b^2) + 48 A M r_+(r-M) 
\nonumber \\
&+& \Delta + 4M r + 4M^2\frac{r + r_+}{r - r_-},
\\
d_1(r) &=& 36 A^2(3\Delta^2 - 4b^4) - 72 A M r_+(r-M) - a^2,
\nonumber \\
& & 
\\
d_2(r) &=& 36 A^2 b^2(3\Delta + 4b^2).
\end{eqnarray}
\end{subequations}
In order to shorten the notation we have also introduced the quantity $b :=
\sqrt{M^2 - a^2} > 0$ (remember that we are excluding the extremal case from
our analysis).

It is simple to verify that $E_A(r,\xi) > 0$ for all $r \geq r_+$ and all
$0\leq \xi\leq 1$ when $A = 0$. Therefore, in the following we assume $A\neq
0$ which implies $d_2(r) > 0$ for all $r\geq r_+$. The strategy of the proof
is to provide a positive lower bound for the quantity
\begin{equation}
f_A(r) := \min\limits_{0\leq \xi\leq 1} E_A(r,\xi)
\label{Eq:EMin}
\end{equation}
for each $r_+\leq r\leq {\cal R}$. For this, we distinguish between the following three cases:
\begin{enumerate}
\item[Case A:] $d_1(r)\geq 0$: In this case the minimum~(\ref{Eq:EMin}) occurs at the poles $\xi = 0$:
\begin{equation}
f_A(r) = E_A(r,0) = d_0(r).
\end{equation}
\item[Case B:] $-2d_2(r) < d_1(r) < 0$: The minimum occurs at $\xi = \xi_* = -d_1(r)/(2d_2(r))$; hence
\begin{equation}
f_A(r) = E_A(r,\xi_*) = d_0(r) - \frac{d_1(r)^2}{4d_2(r)}.
\end{equation}
\item[Case C:] $d_1(r)\leq -2d_2(r)$: The minimum occurs at the equator $\xi = 1$; thus
\begin{equation}
f_A(r) = E_A(r,1) = d_0(r) + d_1(r) + d_2(r).
\end{equation}
\end{enumerate}

We start with case A, for which $E_A(r,\xi)\geq d_0(r)$. Denoting by $A_\pm(r)$
the same function as the one defined in \eq{Eq:Apm} with ${\cal R}$ replaced
with $r$, one has
\begin{equation}
d_0(r) = 144\Delta(r-M)^2 [A_+(r) - A] [A - A_-(r)].
\end{equation}
As one can easily verify, $A_-(r)$ is an increasing function of $r$ while
$A_+(r)$ is a decreasing function of $r$. Therefore, $A_-(r) \leq A_-({\cal
R}) < A < A_+({\cal R})\leq A_+(r)$ for all $r_+\leq r\leq {\cal R}$, which
implies that $d_0(r) > 0$ for all $r_+ < r\leq {\cal R}$. At the horizon,
\begin{equation}
d_0(r_+) = 48 A M r_+ b + 4Mr_+ + 8M^2\frac{r_+}{b},
\end{equation}
which is obviously positive when $A > 0$. When $A < 0$ we use the fact that
\begin{equation}
|A_-({\cal R})| \leq |A_-(r_+)| = \frac{1}{12b}\left( 1 + \frac{M}{b} \right)
\label{Eq:AmBound}
\end{equation}
to conclude that $d_0(r_+) \geq 4M^2 r_+/b > 0$.

Next, we analyze case B for which $0 < -d_1(r) < 2d_2(r)$. This allows us
to estimate
\begin{equation}
\begin{split}
f_A(r) & = d_0(r) + \frac{d_1(r)}{2}\frac{(-d_1(r))}{2d_2(r)}
\\ & \geq d_0(r) + \frac{1}{2} d_1(r)
\geq d_0(r) + \frac{4}{3} d_1(r).
\end{split}
\end{equation}
Explicitly, this yields
\begin{equation}
\begin{split}
f_A(r) \geq &  -144A^2 b^2\left( \Delta + \frac{4}{3} b^2 \right) - 48A M r_+(r-M) \\
& 
 + \Delta + 4M r + 4M^2\frac{r + r_+}{r - r_-} - \frac{4}{3} a^2.
\end{split}
 \end{equation}
For positive $A$ we have the bounds $12A \leq 1/r_+\leq 1/b$ which yields
the estimate
\begin{equation}
\begin{split}
f_A(r) \geq & -\Delta - \frac{4}{3} b^2 - 4M(r-M) + \Delta + 4M r \\
& + 4M^2 - \frac{4}{3} a^2 
  = \frac{16}{3} M^2 > 0.
\end{split}
 \end{equation}
This bound still holds for negative $A$, provided $12 |A_-({\cal R})|
b\leq 1$ which is the case if ${\cal R}\geq M + \sqrt{b(b+2M)}$. If $r_+ <
{\cal R}\leq M + \sqrt{b(b+2M)}$ we use instead the bound~(\ref{Eq:AmBound})
and the fact that $r_+\leq r\leq {\cal R}$ implies $\Delta\leq 2Mb$ to conclude
\begin{widetext}
\begin{eqnarray*}
f_A(r) &\geq& -\left( 1 + \frac{M}{b} \right)\left( \Delta + \frac{4}{3} b^2 \right) 
 + \Delta + 4M r  +\ 4M^2\frac{r + r_+}{r - r_-} - \frac{4}{3} a^2\\
  &\geq& -\frac{M}{b}\Delta + 4M r + \frac{4}{3} M^2 - \frac{4}{3} M b\\
  &\geq& 2M(2r - M) > 0.
\end{eqnarray*}

Finally, in case C the condition $2d_2(r) + d_1(r)\leq 0$ yields
\begin{equation}
108 A^2\Delta(\Delta + b^2)  \leq 108 A^2\left[ (\Delta + b^2)^2 + \frac{1}{3} b^4 \right] \leq 72 AM r_+(r-M) + a^2.
\end{equation}
Therefore,
\begin{eqnarray*}
f_A(r) &=& d_0(r) + d_1(r) + d_2(r)\\
 &=& -36A^2\Delta(\Delta + b^2) - 24 A M r_+(r-M) + r^2 + 2Mr + 4M^2\frac{r + r_+}{r - r_-}\\
 &\geq& -48A Mr_+(r-M) - \frac{a^2}{3} + r^2 + 2Mr + 4M^2\frac{r + r_+}{r - r_-},
\end{eqnarray*}
which is clearly positive when $A < 0$. For $A > 0$ we use the bound $12 A
r_+\leq 1$ and obtain for all $r\geq r_+$
\begin{equation}
f_A(r)\geq -4M(r-M) - \frac{a^2}{3} + r^2 + 2M r + 4M^2 = \Delta + 8M^2 - \frac{4}{3} a^2 > 0.
\end{equation}

This concludes the proof of the theorem.
\qed

\end{widetext}

One can verify that the required hypothesis $12r_+ A_+({\cal R}) \leq 1$ is
always satisfied for ${\cal R}\geq 3M + r_+ = 4M + b$. Although this bound
is not optimal, the condition $A < A_+({\cal R})$ ceases to be sufficient for
small ${\cal R} - r+$, as can be understood from the plots in Fig.~\ref{f3}
which show that in this case, the upper bound on $A$ comes from the equator
(case C in the proof) instead of the poles.

\section{Numerical convergence tests}
\label{App:convergence}

\begin{figure}[t]
 \centering
 \includegraphics[width=0.48\textwidth]{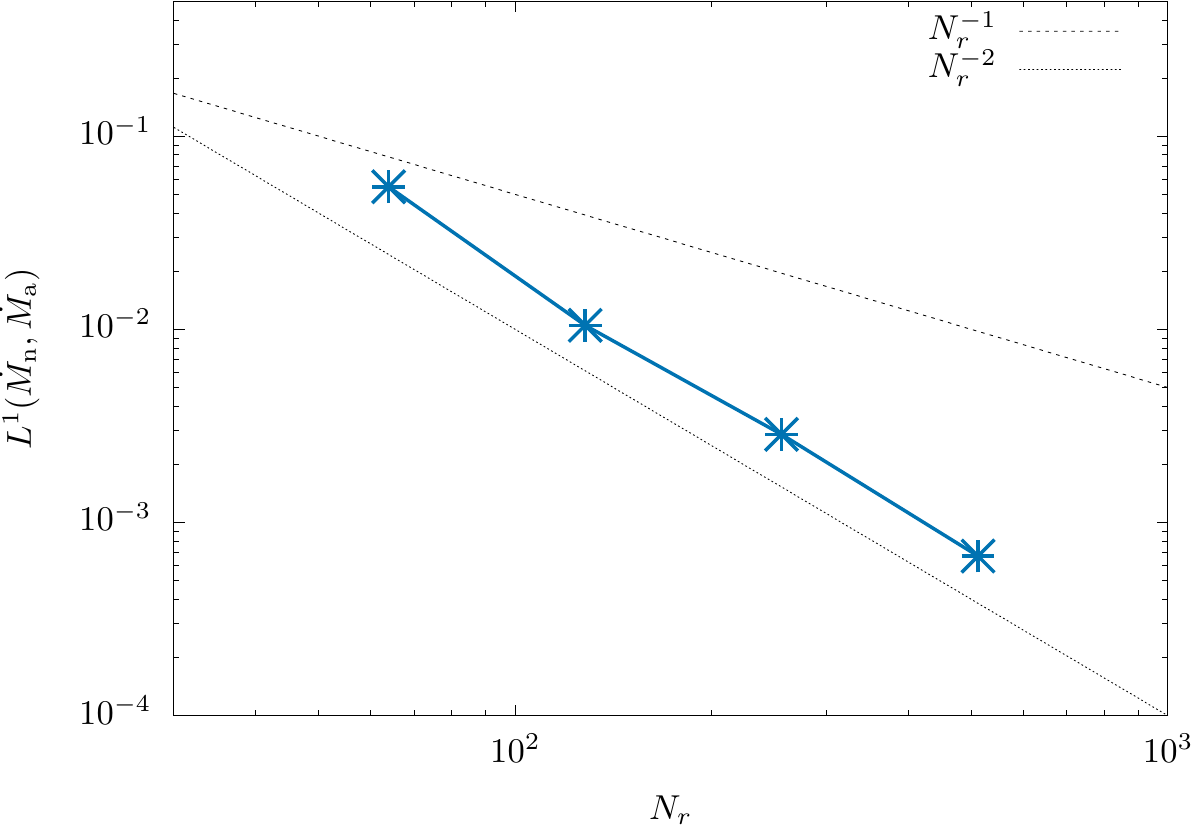}
 \caption{$L^1$-norm of the error in the mass accretion rate, for the benchmark
 test presented in Sec.~\ref{subsec:benchmark}. The test is performed
 using the radial resolutions $N_r= 64, 128, 256, 512$. The black dashed
 lines represent the expected tendency for a first (top) and second order
 (bottom) convergence.}
 \label{fig:conv}
\end{figure}

\begin{figure}[t]
 \centering
 \includegraphics[width=0.48\textwidth]{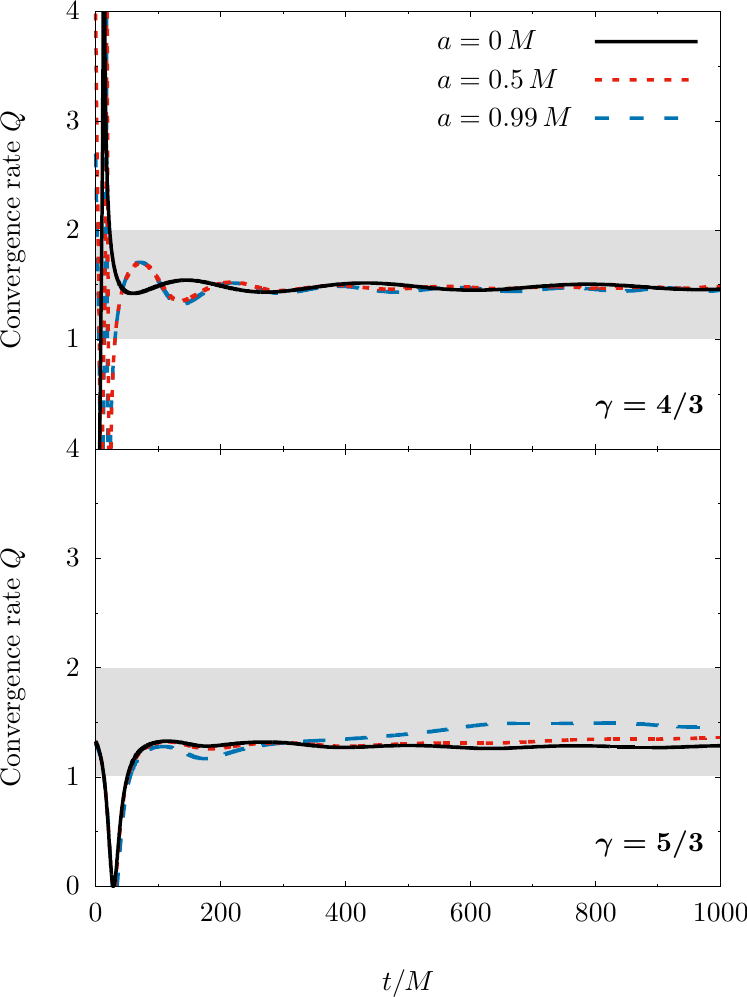}
 \caption{Self-convergence tests of the polytropic fluid simulations,
 for both values of $\gamma$: 4/3 (top panel) and 5/3 (bottom panel). In
 this figure we show the evolution in time of the convergence rate $Q$ for
 three different values of the spin parameter $a$. The
 gray stripe shows the expected convergence zone (given the
 numerical methods used in \textsc{aztekas}).
 }
   \label{fig:self}
\end{figure}

In this appendix we present the convergence and self-convergence tests that are
necessary to validate our numerical results. 

For the benchmark test presented in Sec.~\ref{subsec:benchmark} we compute,
for each resolution studied, the relative error between the numerical and
analytic values of the mass accretion rate, once the steady state has been
reached. In Fig.~\ref{fig:conv} we present the results of these values as a
function of the radial resolution $N_r$, from which we obtain  second order
convergence, as expected for smooth solutions considering the numerical
methods used in {\sc aztekas}.

On the other hand, in order to validate the numerical results of the polytropic
fluid simulations reported in Sec.~\ref{subsec:poly}, we perform a series
of self-convergence tests in which, using three different consecutive
resolutions, we compute the convergence rate of the solution. We carry out
the simulations using resolutions $R_1 =64\times64$, $R_2 = 128\times128$, and
$R_3 = 256\times256$, for each studied value of the adiabatic index $\gamma$
and three different values of the spin parameter $a/M=0,\, 0.5,\, 0.99$.

In Fig.~\ref{fig:self} we show the evolution in time of the convergence rate
$Q$, which is computed as
\begin{equation}
   2^Q = \frac{|\dot{M}_1 - \dot{M}_2|}{|\dot{M}_2 - \dot{M}_3|},
\end{equation}
where $\dot{M}_1$, $\dot{M}_2$ and $\dot{M}_3$ are the values of the mass
accretion rate as obtained from resolutions $R_1$, $R_2$, and $R_3$,
respectively. As can be seen from this figure, the time evolution of
the simulations' convergence rate rapidly becomes confined within the
gray stripe. Note that in this case, since we have set free-outflow
boundary conditions for the velocity field, the simulations develop sharp
global oscillations throughout the domain during their evolution, causing
the convergence rate to be less than second order, which would have been
otherwise expected since we obtain a smooth final steady state solution.

\bibliographystyle{unsrt}
\bibliography{./article}

\end{document}